\documentclass[english, letterpaper, 11pt]{article}

\usepackage{fourier}
\usepackage[T1]{fontenc}

\usepackage[margin=1in]{geometry}
\usepackage{amssymb}
\usepackage{amsmath}
\usepackage{amsmath} 
\usepackage{mathrsfs}
\usepackage{xfrac}
\usepackage{bbm} 
\usepackage{scrextend} 
\usepackage{natbib}
\usepackage{multirow} 
\usepackage{graphicx} 
\usepackage{color}
\usepackage[table,xcdraw]{xcolor}
\usepackage{booktabs} 
\usepackage[labelsep=period]{caption} 
\usepackage[normalem]{ulem} 
\usepackage[hang,flushmargin]{footmisc}
\usepackage[hidelinks]{hyperref}
\usepackage{rotating} 
\usepackage{rotating} 
\usepackage{scalefnt} 
\usepackage{setspace} 
\usepackage{soul} 
\usepackage{enumitem}

\allowdisplaybreaks

\usepackage{authblk} 

\numberwithin{equation}{section}
\newsavebox{\overlongequation}

\begin{document}
	
\title{\Huge Accounting for Cross-Location Technological Heterogeneity in the Measurement of Operations Efficiency and Productivity\thanks{\textbf{Acknowledgments}: Emir Malikov would like to acknowledge financial support from the Troesh Center for Entrepreneurship and Innovation at UNLV. \ \textbf{Correspondence}: Emir Malikov, Lee Business School, University of Nevada, Las Vegas, Las Vegas, NV 89154-6005. Email: emir.malikov@unlv.edu }}
\author[1]{\sc \vspace{0.2cm} Emir Malikov}
\author[2]{\sc Jingfang Zhang}  
\author[3]{\\ \sc Shunan Zhao}
\author[4,5]{\sc Subal C.~Kumbhakar}
\affil[1]{\small Lee Business School, University of Nevada, Las Vegas, Las Vegas, NV, USA } 
\affil[2]{\small Auburn University, Auburn, AL, USA}	
\affil[3]{\small School of Business Administration, Oakland University, Rochester, MI, USA}	
\affil[4]{\small Binghamton University, Binghamton, NY, USA}	
\affil[5]{\small University of Stavanger Business School, Stavanger, Norway}	

\date{\small September 20, 2021}
\maketitle
	
\begin{abstract}
\noindent Motivated by the long-standing interest in understanding the role of location for firm performance, this paper provides a semiparametric methodology to accommodate locational heterogeneity in production analysis. Our approach is novel in that we explicitly model spatial variation in parameters in the production-function estimation. We accomplish this by allowing both the input-elasticity and productivity parameters to be unknown functions of the firm’s geographic location and estimate them via local kernel methods. This allows the production technology to vary across space, thereby accommodating neighborhood influences on firm production. In doing so, we are also able to examine the role of cross-location differences in explaining the variation in operational productivity among firms. Our model is superior to the alternative spatial production-function formulations because it (i) explicitly estimates the cross-locational variation in production functions, (ii) is readily reconcilable with the conventional production axioms and, more importantly, (iii) can be identified from the data by building on the popular proxy-variable methods, which we extend to incorporate locational heterogeneity. Using our methodology, we study China's chemicals manufacturing industry and find that differences in technology (as opposed to in idiosyncratic firm heterogeneity) are the main source of the cross-location differential in total productivity in this industry.
		
\vspace{\baselineskip}
		
 
\end{abstract} 
	
\onehalfspacing
\thispagestyle{empty} \addtocounter{page}{0}
\clearpage
	
	
\section{Introduction}
\label{sec:introduction}

It is well-documented in management, economics as well as operations research that businesses, even in narrowly defined industries, are quite different from one another in terms of productivity. These cross-firm productivity differentials are large, persistent and ubiquitous \citep[see][]{syverson2011}. Research on this phenomenon is therefore unsurprisingly vast and includes attempts to explain it from the perspective of firms' heterogeneous behaviors in research and development \citep[e.g.,][]{griffithetal2004}, corporate operational strategies \citep{smithreece1999}, ability of the managerial teams \citep{demerjianetal2012}, ownership structure \citep{ehrlichetal1994}, employee training and education \citep{moretti2004}, allocation efficiency \citep{songetal2011}, participation in globalization \citep{grossmanhelpman2015} and many others. In most such studies, a common production function/technology is typically assumed for all firms within the industry, and the differences in { operations performance of firms} are {confined to} variation in the {``total factor productivity,''} the Solow residual {\citep{solow1957}}.\footnote{A few studies alternatively specify an ``augmented'' production function which, besides the traditional inputs, also admits various firm-specific shifters such as the productivity-modifying factors mentioned above. {But such studies continue to assume that the same technology frontier applies to all firms.}}

In this paper, we approach the {heterogeneity in firm performance} from a novel perspective in that we explicitly acknowledge the existence of locational effects on the {operations technology of firms} and their underlying productivity. {We} allow the firm-level production function to vary across space, thereby {accommodating potential neighborhood influences on firm production.} In doing so, we are able to examine the role of locational heterogeneity for cross-firm differences in {operations} performance{/efficiency}.
 
\begin{figure}[t]
	\centering
	\includegraphics[scale=0.26]{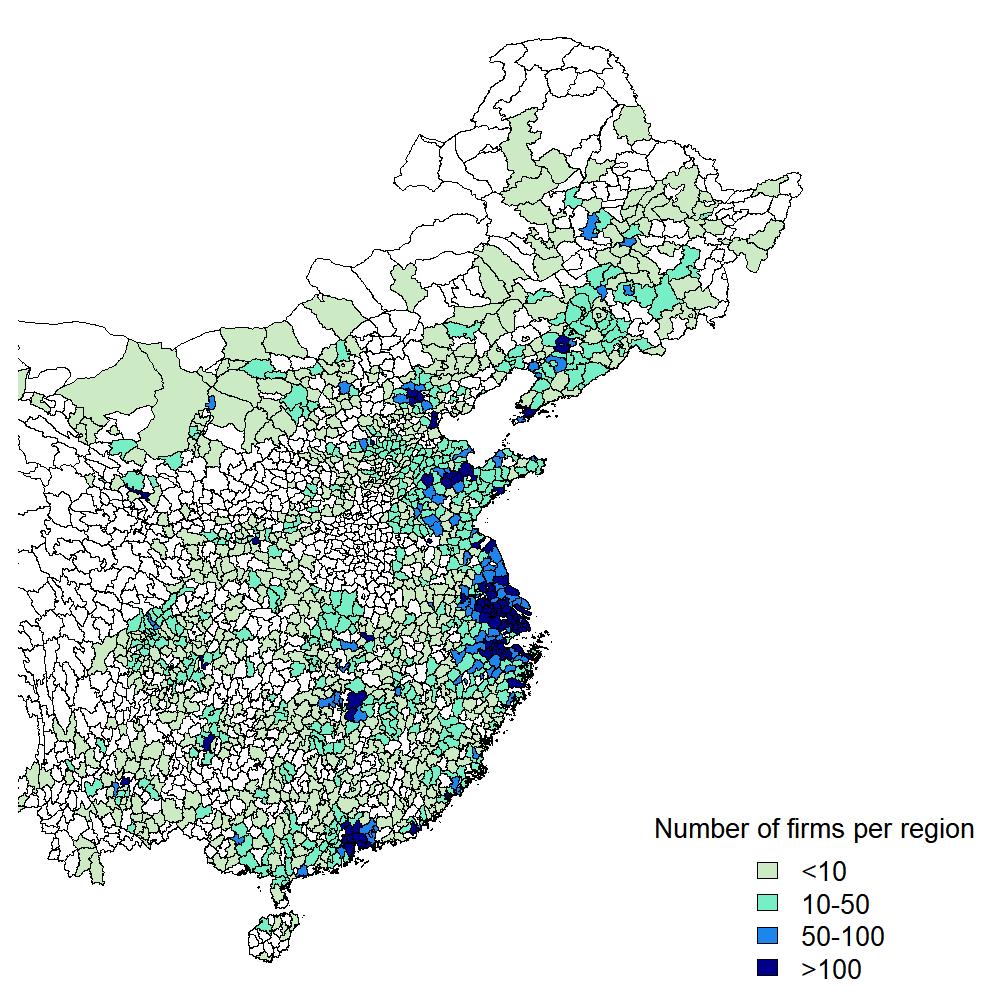}
	\caption{Spatial Distribution of Manufacturers of Chemicals in China, 2004--2006} \label{fig:number}
\end{figure}

{A firm's location is important for its operations technology. For example, \citet{ketokivi2017locate} show that hospital location is significantly related to {its} performance and {that} a hospital’s choice of strategy can {help} moderate the effect of location {through} the interplay of {local} environmental factors with organizational strategy. As shown in Figure \ref{fig:number}, chemical enterprises in China, the focus of empirical analysis in this paper, are widely {(and unevenly)} distributed {across} space. {Given the sheer size of the country (it is the third largest by area), it is implausible that, even after controlling for firm heterogeneity, all these businesses operate using} the same production technology. Organizations in all industries\textemdash not only hospitals and chemical manufacturers\textemdash develop strategies to respond to {local environment} and {the associated} competitive challenges, and those strategies drive operational decisions regarding investments in new or updated technologies}.

{Theoretically,} there are many reasons to believe that the production technology is location-specific. First, exogenous {local} endowments and institutional environments, such as {laws, regulations and local supply chains}, play a key role in determining firm performance. {The location of firms} {determines key linkages between {the} production, market, supply chain and product development \citep{goldsteinetal2002}}. If we look at the global distribution of the supply chains of many products, the product development and design is usually conducted in developed countries such as the U.S. and European countries, while the manufacturing and assembly process is performed in East Asian countries such as China and Vietnam. This spatial distribution largely reflects the endowment differences in factors of production (e.g., skilled vs.~unskilled labor) and the consequent relative input price differentials across countries. 
Analogously, take the heterogeneity in endowment and institutions across different locations \textit{within} a country. There are many more world's leading universities on the East and West Coasts of the U.S.~than in the middle of the country, and they provide thousands of talented graduates each year to the regional development, bolstering growth in flagship industries such as banking and high-tech in those locations. In China, which our empirical application focuses on, networking and political connections are, anecdotally, the key factors for the success of a business in the Northeast regions, whereas the economy on the Southeast Coast is more market-oriented. Furthermore, there are many broadly defined special economics zones (SEZs) in China, which all are characterized by a small designated geographical area, local management, unique benefits and separate customs and administrative procedures \citep[see][]{craneetal2018}. According to a report from the China Development Bank, in 2014, there were 6 SEZs, 14 open coastal cities, 4 pilot free-trade areas and 5 financial reform pilot areas. There were also 31 bonded areas, 114 national high-tech development parks, 164 national agricultural technology parks, 85 national eco-industrial parks, 55 national eco-civilization demonstration areas and 283 national modern agriculture demonstration areas. They spread widely in China and support various economic functions, giving rise to locational heterogeneity in the country's production.
 
Second, most industries are geographically concentrated in general, whereby firms in the same or related industries tend to spatially cluster, benefiting from agglomeration economies reflected, among other things, in their production technologies that bring about localized \textit{aggregate} increasing returns. Ever since \citet{marshall1920} popularized these ideas, researchers have shown that industry concentration is too great to be explained solely by the differences in exogenous locational factors and that there are at least three behavioral micro-foundations for  agglomeration: benefits from labor market pooling/sharing, efficiency gains from the collocation of industries with input-output relationships that improves the quality of matches and technology spillovers  \citep[see][]{ellison_glaeser1999,duranton2004,ellisonetal2010,singh_marx2013}. The key idea of agglomeration economies is that geographic proximity reduces the transport costs of goods, people and, perhaps more importantly, ideas. While it is more intuitive that the movement of goods and people is hindered by spatial distance, the empirical evidence from prior studies shows that technology spillovers are also highly localized because knowledge transfers require interaction that proximity facilitates \citep[see][]{almeida_Kogut1999,alcacer_Chung2007,singh_marx2013}. Therefore, {owing to the role of local neighborhood influences,} firms that produce the same/similar products but are located in regions with different industry concentration levels are expected to enjoy different agglomeration effects on their {operations}. 

Because location is an important factor affecting firm performance, previous empirical studies heavily rely on spatial econometrics to examine the locational/spatial effects on production. Oftentimes, spatially-weighted averages of other firms' outputs and inputs are included as additional regressors in spatial autoregressive (SAR) production-function models \citep[e.g.,][]{glassetal2016,glassetal2020,glassetal2020b, vidoli_Canello2016,serpakrishnan2018, glassetal2019,kutluetal2020,houetal2020}. The appropriateness of such {a conceptualization of firm-level production functions in the presence of locational influences} however remains unclear because {these SAR specifications are difficult to reconcile with the theory of firm.} For instance, the reduced form of such models effectively implies substitutability of the firm's inputs with those of its peers and does not rule out the possibility of the firm's output increasing when the neighboring firms use more inputs even if the firm itself keeps own inputs fixed { and the productivity remains the same. Further, these models continue to implausibly assume that all firms use the same production technology no matter their location. The practical implementation of SAR production-function models is, perhaps, even more problematic: (\textsl{i}) they imply additional, highly nonlinear parameter restrictions necessary to ensure that the conventional production axioms are not violated, and (\textsl{ii}) they are likely unidentifiable from the data given the inapplicability of available proxy-variable estimators and the pervasive lack of valid external instruments at the firm level. We discuss this in detail in Appendix \ref{sec:appx_sar}.}\footnote{But we should note that studies of the nexus between location/geography and firm performance in operations research and management are not all confined to the production theory paradigm; e.g., see \citet{bannisterstolp1995}, \citet{goldsteinetal2002}, \citet{kalnins2004,kalnins2006}, \citet{dahlsorenson2012} and \citet{kulchina2016}.}

In this paper, we consider a semiparametric production function in which both the {input-to-output transformation technology} and  productivity are location-specific. Concretely, using the location information for firms, we let the {input-elasticity} and productivity-process parameters be nonparametric functions of the firm's {geographic location (latitude and longitude)} and estimate these unknown functions via kernel methods. Our methodology captures the cross-firm spatial {influences} through local smoothing, whereby the production technology for each location is calculated as the geographically weighted average of the input-output \textit{relationships} for firms in the nearby locations with larger weights assigned to the firms that are more spatially proximate. This is fundamentally different from the SAR production-function models that formulate {neighborhood influences}  using spatially-weighed averages of the output/inputs \textit{quantities} while keeping the production technology the same for all firms. Consistent with the agglomeration literature, our approach implies that learning and knowledge spillovers are localized and that their chances/intensity diminish with distance. Importantly, by utilizing the data-driven selection of smoothing parameters that regulate spatial weighting of neighboring firms in kernel smoothing, we avoid the need to rely on \textit{ad hoc} specifications of the weighting schemes and spatial radii {of neighborhood influences like the} traditional SAR models do. It also allows us to be agnostic about the channels through which {firm location affects its production,} and our methodology inclusively captures all possible mechanisms of agglomeration economies.  

{Our conceptualization of spatial influences by means of locationally-varying parameters is akin to the idea of ``geographically weighted regressions'' (GWR) introduced and popularized in the field of geography by \citet{bfc1996}; also see \citet{fbc-book} and many references therein. Just like ours, the GWR technique aims to model processes that are not constant over space but exhibit local variations and do so using a varying-coefficient specification estimated via kernel smoothing over locations. However, the principal\textemdash and non-trivial\textemdash distinction of our methodology from the GWR approach is in its emphasis on \textit{identification} of the spatially varying relationship. Concretely, for consistency and asymptotic unbiasedness the GWR methods rely on the assumption that (non-spatial) regressors in the relationship of interest are mean-orthogonal to the stochastic disturbance which rules out the presence of correlated unobservables as well as the potential simultaneity of regressors and the outcome variable for reasons other than spatial autoregression. The latter two are, however, more the rule rather than the exception for economic relations, which are affected by behavioral choices, including the firm-level production function. 
Recovering the data generating process underlying the firm's production operations from observational data (i.e., its identification) requires tackling the correlation between regressors and the error term that the GWR cannot handle, making it unable to consistently estimate the production technology and firm productivity. This is precisely our focus.\footnote{In effect, our methodology constitutes a generalization of the GWR technique to accommodate endogenous regressors in the context of production-function estimation.}}

The identification of production functions in general, let alone with locational heterogeneity, is not trivial due to the endogeneity issue whereby the firm's input choices are correlated with its productivity. Complexity stems from the latency of firm productivity. Due to rather unsatisfactory performance of the conventional approaches to identification of production functions, such as fixed effects estimation or instrumentation using prices, there is a growing literature targeted at solving endogeneity using a proxy-variable approach \citep[e.g., see][]{op1996,lp2003,acf2015,gnr2013} which has gained wide popularity among empiricists.

To identify the locationally-varying production functions, we develop a semiparametric proxy-variable estimator that accommodates locational heterogeneity across firms. To this end, we build upon  \citet{gnr2013} whose framework we extend to incorporate spatial information about the firms in a semiparametric fashion. More specifically, we make use of the structural link between the production function (of the known varying-coefficient functional form) and the optimality condition for a flexible input derived from the firm's static expected profit maximization problem. We propose a two-step estimation procedure and, to approximate the unknown functional coefficients, employ  local-constant kernel fitting. Based on the estimated location-specific production functions, we further propose a locational productivity differential decomposition to break down the cross-region production differences that cannot be explained by input usage (i.e., the differential in ``total productivity'' of firms across locations) into the contributions attributable to differences in available production technologies  and to differences in {total-factor operations efficiency of firms.} 

We apply our model to study locationally heterogeneous production technology among Chinese manufacturing firms in the chemical industry in 2002--2004. Based on the results of the data-driven cross-validation as well as formal statistical tests, the empirical evidence provides strong support to the importance and relevance of location for production. Qualitatively, we find that both technology and {firm} productivity vary significantly across regions. Firms are more likely to exhibit higher {(internal)} returns to scale in regions of agglomeration. However, the connection between {firm} productivity and industry concentration {across space} is unclear. The decomposition analysis reveals that differences in {\textit{technology} (as opposed to in idiosyncratic firm heterogeneity)} are the main source of cross-location total productivity differentials, on average {accounting for 2/3} of the differential. 
 
To summarize, our contribution is as follows. We propose a semiparametric methodology to accommodates locational heterogeneity in the production-function estimation while maintaining the standard structural assumptions about firm production. {Unlike the available SAR-type alternatives, our model explicitly estimates the cross-locational variation in production technology.} To operationalize our methodology, we extend the widely-used proxy-variable identification methods to incorporate {firm location}. Our model as well as the proposed decomposition method for disentangling the effects of location on {firm} productivity from those on technological input-output relationship should provide a valuable addition to the toolkit of {empiricists} interested in studying agglomeration economies and technology spillovers. {In the context of operations management in particular, our methodology will be most useful for empirical studies focused on the analysis of operations efficiency/productivity and} {its ``determinants;'' \citep[e.g.,][are just a few recent examples of such analyses]{ross2004analysis,berenguer2016disentangling,jola2016effect,lam2016impact}. {In the case of multi-input production, the ``total factor productivity'' is among the most popular comprehensive measures of operations efficiency/productivity of the firm,} and our paper shows how to measure the latter robustly when production relationships are not constant over space {and are subject to neighborhood influences. This is particularly interesting because the effects of location, supply chain integration and agglomeration on firm performance have recently attracted much attention among researchers in operations management} \citep[e.g.,][]{goldsteinetal2002,ketokivi2017locate,flynn2010impact}.}

The rest of the paper is organized as follows. Section \ref{sec:model} describes the model of firm-level production {exhibiting} locational heterogeneity. We describe our identification and estimation strategy in Section \ref{sec:identification_estimation}. We provide the locational productivity differential decomposition in Section \ref{sec:decomposition_level}. The empirical application is presented in Section \ref{sec:application}. Section \ref{sec:conclusion} concludes. Supplementary materials are relegated to the Appendix.


\section{Locational Heterogeneity in Production}
\label{sec:model}

Consider the production process of a firm $i$ ($i=1,\dots,n$) in the time period $t$ ($t=1,\dots,T$) in which physical capital $K_{it} {\in\Re_{+}}$, labor $L_{it} {\in\Re_{+}}$ and an intermediate input such as materials $M_{it}{\in\Re_{+}}$ are  transformed into the output $Y_{it}{\in\Re_{+}}$ via a production function given the (unobserved) firm productivity. Also, let $S_{i}$ be the (fixed) location of firm $i$, with the obvious choice being $S_{i}=(\text{lat}_i,\text{long}_i)'$, where $\text{lat}_i$ and $\text{long}_i$ are the latitude and longitude coordinates of the firm's location. Then, the locationally varying production function is
\begin{equation}\label{eq:prodfn_hicks_f}
Y_{it} = F_{|S_{i}}(K_{it},L_{it},M_{it})\exp\left\{\omega_{it}\right\}\exp\left\{\eta_{it}\right\} ,
\end{equation}
where $F_{|S_{i}}(\cdot)$ is the firm's location-specific production {function} that {varies} over space (as captured by $S_i$) to accommodate locational heterogeneity in production technology, $\omega_{it}$ is the firm's persistent Hicks-neutral {total factor productivity capturing its operations efficiency}, and $\eta_{it}$ is a random transitory shock. Note that, so long as the firm's location is fixed, $\omega_{it}$ that persist for the same firm $i$, by implication, then also has the evolution process that is specific to this firm's location $S_i$; we expand on this below.

As in \citet{gnr2013}, \citet{malikovetal2020} and \citet{malikovzhao2021}, physical capital $K_{it}$ and labor $L_{it}$ are said to be subject to adjustment frictions (e.g., time-to-install, hiring/training costs), and the firm optimizes them dynamically at time $t-1$ rendering these predetermined inputs quasi-fixed at time $t$. Materials $M_{it}$ is a freely varying (flexible) input and is determined by the firm statically at time $t$. Thus, both $K_{it}$ and $L_{it}$ are the state variables with dynamic implications and follow their respective deterministic laws of motion: 
\begin{equation}\label{eq:k_lawofmotion}
	K_{it}=I_{it-1}+(1-\delta)K_{it-1}\quad \text{and}\quad L_{it}=H_{it-1}+L_{it-1} ,
\end{equation}
where $I_{it}$, $H_{it}$ and $\delta$ are the gross investment, net hiring and the depreciation rate, respectively. 

Following the convention, we assume that the risk-neutral firm maximizes a discounted stream of expected life-time profits in perfectly competitive output and factor markets subject to the state variables and expectations about the market structure variables including prices that are common to all firms.\footnote{{
We use the perfect competition and homogeneous price assumption mainly for two reasons: (\textsl{i}) it is the most widely used assumption in the literature on structural identification of production function and productivity, and (\textsl{ii}) this assumption has been {repeatedly} used {when} studying the same data as ours \citep[e.g.,][]{baltagietal2016,malikovetal2020}. Relaxing the perfect competition assumption is possible but non-trivial, and it requires additional assumptions {about} the output demand \citep[e.g.,][]{deLoecker2011product} and/or extra information on firm-specific output prices that are usually not available for manufacturing data \citep[e.g.,][]{deloeckeretal2016} {or imposing \textit{ex ante} structure on the returns to scale \citep[see][]{flynnetal2019}.} It is still a subject of ongoing research. Given {the emphasis of our contribution on incorporating \textit{technological} heterogeneity (associated with firm location, in our case) in the measurement of firm productivity, we opt} to keep all other aspects of {modeling} consistent with {the convention in the literature to ensure meaningful comparability with most available methodologies.}}} Also, for convenience, we denote $\mathcal{I}_{it}$ to be the information set available to the firm $i$ for making the period $t$ production decisions.

In line with the proxy variable literature, we model firm productivity $\omega_{it}$ as a first-order Markov process which we, however, endogenize \`a la \citet{dj2013} and \citet{deloecker2013} by incorporating productivity-enhancing and ``learning'' activities of the firm. To keep our model as general as possible, we denote all such activities via a generic variable $G_{it}$ which, depending on the empirical application of interest, may measure the firm's R\&D expenditures, foreign investments, export status/intensity, etc.\footnote{A scalar variable $G_{it}$ can obviously be replaced with a vector of such variables.} Thus, $\omega_{it}$ evolves according to a location-inhomogeneous controlled first-order Markov processes with transition probability $\mathcal{P}^{\omega}_{|S_i}(\omega_{it}|\omega_{it-1},G_{it-1})$. This implies the following location-specific mean regression for firm productivity:
\begin{equation}\label{eq:productivity_hicks_law}
\omega_{it}=h_{|S_i}\left(\omega_{it-1},G_{it-1}\right)+\zeta_{it},
\end{equation}
where $h_{|S_i}(\cdot)$ is the location-specific conditional mean function of $\omega_{it}$, and $\zeta_{it}$ is a random innovation unanticipated by the firm at period $t-1$ and normalized to zero mean: 
$ \mathbb{E}\left[\zeta_{it} | \mathcal{I}_{it-1}\right]=\mathbb{E}\left[\zeta_{it}\right]=0$.

The evolution process in \eqref{eq:productivity_hicks_law} implicitly assumes that productivity-enhancing activities and learning take place with a delay which is why the dependence of $\omega_{it}$ on a control $G_{it}$ is lagged implying that the improvements in firm productivity take a period to materialize. Further, in $\mathbb{E}[\zeta_{it} |\ \mathcal{I}_{it-1}]=0$ we assume that, due to adjustment costs, firms do not experience changes in their productivity-enhancing investments in light of expected \textit{future} productivity innovations. Since the innovation $\zeta_{it}$ represents inherent uncertainty about productivity evolution as well as the uncertainty about the success of productivity-enhancing activities, the firm relies on its knowledge of the \textit{contemporaneous} productivity  $\omega_{it-1}$ when choosing the level of $G_{it-1}$ in period $t-1$ while being unable to anticipate $\zeta_{it}$. These structural timing assumptions are commonly made in models with controlled productivity processes \citep[e.g.,][]{vanbiesebroeck2005, dj2013,dj2018, deloecker2013, malikovetal2020,malikovzhao2021} and are needed to identify the within-firm productivity-improving learning effects. 

We now formalize the firm's optimization problem in line with the above discussion. Under risk neutrality, the firm's optimal choice of freely varying input $M_{it}$ is described by the (static) restricted expected profit-maximization problem subject to the already optimal dynamic choice of quasi-fixed inputs:
\begin{align}\label{eq:profitmax}
\max_{M_{it}}\ P_{t}^Y F_{|S_i}(K_{it},L_{it},M_{it})\exp\left\{\omega_{it}\right\}\theta  - P_{t}^M M_{it} ,
\end{align}
where $P_{t}^Y$ and $P_{t}^M$ are respectively the output and material prices that, given the perfect competition assumption, need not vary across firms; and $\theta\equiv\mathbb{E}[\exp\{\eta_{it}\}|\ \mathcal{I}_{it}]$. The first-order condition corresponding to this optimization yields the firm's conditional demand for $M_{it}$.

Building on \citeauthor{dj2018}'s (2013, 2018) treatment of productivity-enhancing R\&D investments (a potential choice of $G_{it}$ in our framework) as a contemporaneous decision, we describe the firm's dynamic optimization problem by the following Bellman equation:
\begin{align}\label{eq:bellman}
	\mathbb{V}_{t}\big(\Xi_{it}\big) = &\max_{I_{it},H_{it},G_{it}} \Big\{ \Pi_{t|S_i}(\Xi_{it}) -
	\text{C}^{I}_{t}(I_{it}) -
	\text{C}^{H}_{t}(H_{it})-
	\text{C}^{G}_{t}(G_{it}) + \mathbb{E}\Big[\mathbb{V}_{t+1}\big(\Xi_{it+1}\big) \Big| \Xi_{it},I_{it},H_{it},G_{it}\Big]\, \Big\} ,
\end{align}
where $\Xi_{it}=(K_{it},L_{it},\omega_{it})'\in \mathcal{I}_{it}$ are the state variables;\footnote{The firm's location $S_i$ is suppressed in the list of state variables due to its time-invariance.} $\Pi_{t|S_i}(\Xi_{it})$ is the restricted profit function derived as a value function corresponding to the static problem in \eqref{eq:profitmax}; and $\text{C}^{\kappa}_t(\cdot)$ is the cost function for capital ($\kappa=I$), labor ($\kappa=H$) and productivity-enhancing activities ($\kappa=G$).\footnote{The assumption of separability of cost functions is unimportant, and one can reformulate \eqref{eq:bellman} using one $\text{C}_{t}(I_{it},H_{it},G_{it})$ for all dynamic production variables.} In the above dynamic problem, the level of productivity-enhancing activities $G_{it+1}$ is chosen in time period $t+1$ unlike the amounts of dynamic inputs $K_{it+1}$ and $L_{it+1}$ that are chosen by the firm in time period $t$ (via $I_{it}$ and $H_{it}$, respectively). Solving \eqref{eq:bellman} for $I_{it}$, $H_{it}$ and $G_{it}$ yields their respective optimal policy functions.

{An important assumption of our structural model of firm production in the presence of locational heterogeneity is that firm location $S_i$ is both fixed and exogenous. However, the identification of locational heterogeneity in production may be complicated by the potentially endogenous spatial sorting problem, whereby more productive firms might \textit{ex ante} sort into the what-then-become high productivity locations. Under this scenario, when we compare firm productivity and technology across locations, we may mistakenly attribute gradients therein to the locational effects such as agglomeration and neighborhood influences, while in actuality it may be merely reflecting the underlying propensity of all firms in a given location to be more productive \textit{a priori}. While there has recently been notable progress in formalizing and understanding these coincident phenomena theoretically \citep[e.g.,][]{behrensetal2014,gaubert2018}, disentangling firm sorting and spatial agglomeration remains a non-trivial task empirically.\footnote{Urban economics literature also distinguishes the third endogenous process usually referred to as the ``selection'' which differs from sorting  in that it occurs \textit{ex post} after the firms had self-sorted into locations and which determines their continuing survival. We abstract away from this low-productivity-driven attrition issue in the light of the growing empirical evidence suggesting that it explains none of spatial productivity differences which, in contrast, are mainly driven by agglomeration economies \citep[see][]{combesetal2012}. Relatedly, the firm attrition out of the sample has also become commonly accepted as a practical non-issue in the productivity literature so long as the data are kept unbalanced. For instance, \citet[][p.324]{lp2003} write: ``The original work by Olley and Pakes devoted significant effort to highlighting the importance of not using an artificially balanced sample (and the selection issues that arise with the balanced sample). They also show once they move to the unbalanced panel, their selection correction does not change their results.''} 
However, by including the firm’s own lagged productivity in the autoregressive $\omega_{it}$ process in \eqref{eq:productivity_hicks_law}, we are able (at least to some extent) to account for this potential self-sorting because sorting into locations is heavily influenced by the firm's own productivity (oftentimes stylized as the ``talent'' or ``efficiency'' in theoretical models). That is, the locational heterogeneity in firm productivity and technology in our model is measured after partialling out the contribution of its own past productivity. Incidentally, \citet{deloecker2013} argues the same in the context of productivity effects of exporting and self-selection of exporters.}


\section{Methodology}
\label{sec:identification_estimation}

This section describes our strategy for (structural) identification and estimation of the firm's location-specific production technology and unobserved productivity.

Following the popular practice in the literature \citep[e.g., see][]{op1996,lp2003, dj2013,acf2015,collarddeloecker2015, koningsvanormelingen2015}, we assume the Cobb-Douglas specification for the production function which we adapt to allow for potential locational heterogeneity in production. We do so in a semiparametric fashion as follows:
\begin{equation}\label{eq:prodfn_hicks_cd}
\ln F_{|S_i}(\cdot) = \beta_K(S_i) k_{it}+\beta_L(S_i)l_{it}+\beta_M(S_i)m_{it} ,
\end{equation}
where the lower-case variables denote the logs of the corresponding upper-case variables, and the input elasticity functions $[\beta_K(\cdot),\beta_L(\cdot),\beta_M(\cdot)]'$ are unspecified {smooth} functions of the firm's location $S_i$. {The local smoothness feature of the production {relationship}, including both {the} input elasticities and persistent productivity {process below}, captures the effects of technology spillovers and agglomeration economies} {that give rise to local neighborhood influences}. Our methodology can also adopt more 
flexible specifications such as the log-quadratic translog, which provides a natural extension of the log-linear Cobb-Douglas form. See Appendix \ref{sec:appx_tl} for the details on this extension.

Before proceeding further, we formalize the location-specific autoregressive conditional mean function of $\omega_{it}$ in its evolution process  \eqref{eq:productivity_hicks_law}. Following \citet{dj2013,dj2019}, \citet{acf2015}, \citet{griecoetal2016,griecoetal2019} and many others, we adopt a parsimonious first-order autoregressive specification of the Markovian evolution for productivity but take a step further by assuming a more flexible semiparametric location-specific formulation akin to that for the production technology in \eqref{eq:prodfn_hicks_cd}:
\begin{equation}\label{eq:productivity_hicks_lawsp}
h_{|S_i}(\cdot ) = \rho_0(S_i)+ \rho_1(S_i) \omega_{it-1}+\rho_2(S_i)G_{it-1}.
\end{equation} 


\subsection{Proxy Variable Identification}
\label{sec:identification}

Substituting for $F_{|S_i}(\cdot)$ in the locationally varying production function \eqref{eq:prodfn_hicks_f} using \eqref{eq:prodfn_hicks_cd},  we obtain
\begin{align}
y_{it} &= \beta_K(S_i) k_{it}+\beta_L(S_i)l_{it}+\beta_M(S_i)m_{it} + \omega_{it}+ \eta_{it} \label{eq:prodfn_hicks_cd2} \\
&= \beta_K(S_i) k_{it}+\beta_L(S_i)l_{it}+\beta_M(S_i)m_{it} +  \rho_0(S_i)+ \rho_1(S_i) \omega_{it-1}+\rho_2(S_i)G_{it-1} + \zeta_{it}+ \eta_{it}, \label{eq:prodfn_hicks_cd3}
\end{align}
where we have also used the Markov process for $\omega_{it}$ from \eqref{eq:productivity_hicks_law} combined with \eqref{eq:productivity_hicks_lawsp} in the second line.

Under our structural assumptions, all right-hand-side covariates in \eqref{eq:prodfn_hicks_cd3} are predetermined and weakly exogenous with respect to $\zeta_{it} + \eta_{it}$, except for the freely varying input $m_{it}$ that the firm chooses in time period $t$ conditional on $\omega_{it}$ (among other state variables) thereby making it a function of $\zeta_{it}$. That is, $m_{it}$ is endogenous. 

Prior to finding ways to tackle the endogeneity of $m_{it}$, to consistently estimate \eqref{eq:prodfn_hicks_cd3}, we first need to address the latency of firm productivity $\omega_{it-1}$. A popular solution is a proxy variable approach \`a la \citet{lp2003} whereby latent productivity is controlled for by inverting the firm's conditional demand for an observable static input such as materials. However, such a standard proxy approach generally fails to identify the firm's production function and productivity due to the lack of a valid instrument (from within the production function) for the endogenous $m_{it}$ despite the abundance of predetermined lags of inputs. As recently shown by \citet{gnr2013}, identification cannot be achieved using the standard procedure because \textit{no} exogenous higher-order lag provides excluded relevant variation for $m_{it}$ after conditioning the model on the already included self-instrumenting variables. As a result, the production function remains unidentified in flexible inputs. In order to solve this under-identification problem, \citet{gnr2013} suggest exploiting a structural link between the production function and the firm's (static) first-order condition for the freely varying input. In what follows, we build on this idea which we modify along the lines of \citet{dj2013} and \citet{malikovzhao2021} in explicitly making use of the assumed functional form of production technology.

\textsl{First step}.\textemdash We first focus on the identification of production function in its flexible input $m_{it}$. Specifically, given the technology specification in \eqref{eq:prodfn_hicks_cd}, we seek to identify the material elasticity function $\beta_M(S_i)$. To do so, we consider an equation for the firm's first-order condition for the static optimization in \eqref{eq:profitmax}. The optimality condition with respect to $M_{it}$ in logs is given by (in logs)
\begin{equation}\label{eq:foc_cd}
\ln P_{t}^Y+\beta_K(S_i) k_{it}+\beta_L(S_i)l_{it}+\ln \beta_M(S_i)+[\beta_M(S_i)-1]m_{it} + \omega_{it}+ \ln \theta = \ln P_{t}^M ,
\end{equation}
which can be transformed by subtracting the production function in \eqref{eq:prodfn_hicks_cd2} from it to obtain the following location-specific material share equation:
\begin{equation}\label{eq:fst}
v_{it} = \ln [\beta_M(S_i)\theta] - \eta_{it} ,
\end{equation}
where $v_{it} \equiv \ln \left(P_{t}^M M_{it}\right)-\ln \left( P_{t}^Y Y_{it}\right)$ is the log nominal share of material costs in total revenue, which is observable in the data.

The material share equation in \eqref{eq:fst} is powerful in that it enables us to identify unobservable material elasticity function $\beta_M(S_i)$ using the information about the log material share $v_{it}$. Specifically, we first identify a ``scaled'' material elasticity function $\beta_M(S_i)\times\theta$ using the  moment condition $\mathbb{E}[\eta_{it} | \mathcal{I}_{it}] =\mathbb{E}\left[\eta_{it}|S_i\right]=0$, from where we have that
\begin{equation}\label{eq:fst_ident_theta}
\ln [\beta_M(S_i)\theta] = \mathbb{E}[v_{it}|S_i].
\end{equation}

To identify the material elasticity function $\beta_M(S_i)$ net of constant $\theta$, note that $\theta$ is 
\begin{align}\label{eq:theta}
\theta&\equiv \mathbb{E}\left[\exp\left\{ \eta_{it}\right\} \right]=\mathbb{E}\left[\exp\left\{ \eta_{it}\right\} \right]
=\mathbb{E}\left[\exp\left\{ \mathbb{E}[v_{it}|S_i]-v_{it}\right\} \right],
\end{align}
which allows us to isolate $\beta_M(S_i)$ via
\begin{equation}\label{eq:fst_ident}
\beta_M(S_i) = \exp\left\{ \mathbb{E}[v_{it}|S_i] \right\}/ 
\mathbb{E}\left[\exp\left\{ \mathbb{E}[v_{it}|S_i]-v_{it}\right\} \right].
\end{equation}

By having identified the material elasticity function $\beta_M(S_i)$, we have effectively pinpointed the production technology in the dimension of its endogenous static input thereby effectively circumventing the \citet{gnr2013} critique. This is evident when \eqref{eq:prodfn_hicks_cd3} is rewritten as
\begin{equation}\label{eq:prodfn_hicks_cd4}
y_{it}^* = \beta_K(S_i) k_{it}+\beta_L(S_i)l_{it}+ \rho_0(S_i)+ \rho_1(S_i) \omega_{it-1}+\rho_2(S_i)G_{it-1} + \zeta_{it}+ \eta_{it} ,
\end{equation}
where $y_{it}^*\equiv y_{it} - \beta_M(S_i) m_{it}$ on the left-hand side is already identified/observable and, hence, model in \eqref{eq:prodfn_hicks_cd4} now contains \textit{no} endogenous regressors that need instrumentation.

\textsl{Second step}.\textemdash To identify the rest of the production function, we proxy for latent $\omega_{it-1}$ using the known functional form of the conditional material demand function implied by the static first-order condition in \eqref{eq:foc_cd} which we analytically invert for productivity. Namely, using the inverted (log) material function $\omega_{it}=\ln[ P_{t}^M/P_{t}^Y]-\beta_K(S_i) k_{it}-\beta_L(S_i)l_{it}-\ln [\beta_M(S_i)\theta]+[1-\beta_M(S_i)]m_{it}$ to substitute for $\omega_{it-1}$ in \eqref{eq:prodfn_hicks_cd4}, we get
\begin{align}\label{eq:sst}
y_{it}^* &= \beta_K(S_i) k_{it}+\beta_L(S_i)l_{it}+ \rho_0(S_i)+ \rho_1(S_i) \Big[\nu^*_{it-1}-\beta_K(S_i) k_{it-1}-\beta_L(S_i)l_{it-1}\Big]+\rho_2(S_i)G_{it-1} + \zeta_{it}+ \eta_{it} ,
\end{align}
where $\nu^*_{it-1}=\ln[ P_{t-1}^M/P_{t-1}^Y]-\ln [\beta_M(S_i)\theta]+[1-\beta_M(S_i)]m_{it-1}$ is already identified/observable and predetermined with respect to $\zeta_{it}+ \eta_{it}$.\footnote{{Following the convention {in the} literature \citep[e.g.,][]{op1996,lp2003,acf2015,dj2013,gnr2013}, we assume there is no measurement error in $m_{it}$. However, if the (log) material input is measured with errors, due to {the} reasons such as inventories, subcontracting and outsourcing, it will affect both the first- and the second-step estimation. More specifically, {adjusting} the first step is less problematic if {the measurement error is classical} since $m_{it}$ is in the dependent variable. However, {the second-step} equation \eqref{eq:sst} {will have a new} endogeneity issue due to the measurement error. In such a case, additional {identifying} assumptions are often needed; {see} \citet{hu2020estimating} for an example.}} All regressors in \eqref{eq:sst} are weakly exogenous, and this proxied model is identified based on the moment conditions:
\begin{equation}\label{eq:sst_ident}
\mathbb{E}[ \zeta_{t} + \eta_{t} |\ k_{it},l_{it},k_{it-1},l_{it-1},G_{it-1},\nu^*_{it-1}(m_{it-1}), S_{i} ] = 0.
\end{equation}

With the production technology and the transitory shock ${\eta}_{it}$ successfully identified in the two previous steps, we can readily recover $\omega_{it}$ from \eqref{eq:prodfn_hicks_cd2} via $\omega_{it}=y_{it}- \beta_K(S_i)k_{it}-\beta_L(S_i)l_{it}-\beta_M(S_i)m_{it}-\eta_{it}$.

Our identification methodology is also robust to the \citet{acf2015} critique that focuses on the inability of structural proxy estimators to separably identify the additive production function and productivity proxy. Such an issue normally arises in the wake of perfect functional dependence between freely varying inputs appearing both inside the unknown production function and productivity proxy. Our second-step equation \eqref{eq:sst} does not suffer from such a problem because it contains no (endogenous) variable input on the right-hand side, the corresponding elasticity of which has already been identified from the material share equation in the first step.


\subsection{Semiparametric Estimation}
\label{sec:estimation}

Given the semiparametric varying-coefficient specifications adopted for both the production technology [in \eqref{eq:prodfn_hicks_cd}] and productivity evolution [in \eqref{eq:productivity_hicks_lawsp}], we estimate both the first- and second-step equations \eqref{eq:fst} and \eqref{eq:sst} via \textit{local} least squares. We employ local-constant kernel fitting.

Denote the unknown $\ln [\beta_M(S_i)\theta]$ as some nonparametric function $b_M(S_i)$. Under the assumption that input elasticity functions are smooth and twice continuously differentiable in the neighborhood of $S_i=s$, unknown $b_M(S_i)$ can be locally approximated around $s$ via $b_M(S_i) \approx b_M(s)$ at points $S_i$ close to $s$, $|S_i-s|=o(1)$. Therefore, for locations $S_i$ in the neighborhood of $s$, we can approximate \eqref{eq:fst} by
\begin{equation}\label{eq:fst_est}
v_{it} \approx b_M(s) - \eta_{it} ,
\end{equation}
with the corresponding local-constant kernel estimator of $\ln [\beta_M(s)\theta]$ given by 
\begin{equation}\label{eq:fst_est2_pre}
\widehat{b}_M(s) = \left[\sum_i\sum_t \mathcal{K}_{{h}_1}(S_i,s)\right]^{-1}\sum_i\sum_t \mathcal{K}_{{h}_1}(S_i,s)v_{it},
\end{equation}
where {$\mathcal{K}_{h_1}(S_i,s)$ is a kernel} that weights each observation on the basis of proximity of its $S_i$ value to $s$. 

{To avoid over-smoothing in dense ranges of the support of the data while under-smoothing in
sparse tails, which a ``fixed'' bandwidth parameter is well-known to produce, we employ an ``adaptive'' bandwidth capable of adapting to the local distribution of the data. Specifically, to weight observations, we use an $h_1$-nearest-neighbor bandwidth $R_{h_1}(s)$ defined as the Euclidean distance between the fixed location $s$ and its $h_1$th nearest location among $\{S_i\}$, i.e.,
\begin{equation}\label{eq:knn-bw}
R_{h_1}(s)=\lVert S_{(h_1)}-s\lVert,
\end{equation}
where $S_{(h_1)}$ is the $h_1$th nearest neighbor of $s$. Evidently, $R_{h_1}(s)$ is just the $h_1$th order statistic on the distances $\lVert S_i-s\lVert$. It is $s$-specific and, hence, adapts to data distribution. Correspondingly, the kernel weight function is given by 
\begin{equation}
 \mathcal{K}_{h_1}(S_i,s)=\mathsf{K}\left(\frac{\lVert S_i-s\lVert}{R_{h_1}(s)}\right) ,
\end{equation}
where $\mathsf{K}(\cdot)$ is a (non-negative) smooth kernel function such that $\int \mathsf{K}(\lVert u\lVert)du=1$; we use a second-order Gaussian kernel. 

The key parameter here that controls the degree of smoothing in the first-step estimator \eqref{eq:fst_est2_pre} is the number of nearest neighbors (i.e., locations) $h_1$, which diverges to $\infty$ as $n\to \infty$ but slowly: $h_1/n\to 0$. We select the optimal $h_1$ using the data-driven cross-validation procedure. Also note that, despite the location $S_i$ being multivariate, the parameter $h_1$ is a scalar because it modulates a univariate quantity, namely the distance. Hence, the  bandwidth $R_{h_1}(s)$ is also scalar. That is, unlike in the case of a more standard kernel fitting based on fixed bandwidths when the data are weighted using the product of univariate kernels corresponding to each element in $S_i-s$, the adaptive kernel fitting weights data using a \textit{norm} of the vector $S_i-s$. For this reason, when employing nearest neighbor methods, the elements of smoothing variables  are typically rescaled so that they are all comparable because, when $S_i$ is multivariate, the nearest neighbor ordering is not scale-invariant. In our case however, we do \textit{not} rescale the elements of $S_i$ (i.e., latitude and longitude) because they are already measured on the same scale and the (partial) distances therein have a concrete physical interpretation.

From \eqref{eq:fst_ident},} the first-step estimator of $\beta_M(s)$ is
\begin{equation}\label{eq:fst_est2}
\widehat{\beta}_M(s) = nT\exp\left\{ \widehat{b}_M(s) \right\}\Big/ \sum_i\sum_t\exp\left\{ \widehat{b}_M(s)-v_{it}\right\} .
\end{equation}

We construct $\widehat{y}_{it}^*\equiv y_{it} - \widehat{\beta}_M(S_i) m_{it}$ and $\widehat{\nu}^*_{it-1}=\ln[ P_{t-1}^M/P_{t-1}^Y]-\ln [\widehat{\beta}_M(S_i)\theta]+[1-\widehat{\beta}_M(S_i)]m_{it-1}$ using the first-step local estimates of $\beta_M(S_i)$. Analogous to the first-step estimation, we then locally approximate each unknown parameter function in \eqref{eq:sst} around $S_i=s$ via local-constant approach. Therefore, for locations $S_i$ near $s$, we have
\begin{align}\label{eq:sst_est}
\widehat{y}_{it}^* &\approx \beta_K(s) k_{it}+\beta_L(s)l_{it}+ \rho_0(s)+ \rho_1(s) \Big[\widehat{\nu}^*_{it-1}-\beta_K(s) k_{it-1}-\beta_L(s)l_{it-1}\Big]+\rho_2(s)G_{it-1} + \zeta_{it}+ \eta_{it}.
\end{align}

Denoting all unknown parameters in \eqref{eq:sst_est} collectively as $\Theta(s)=[\beta_K(s),\beta_L(s),\rho_0(s),\rho_1(s),\rho_2(s)]'$, we estimate the second-step equation via locally weighted {nonlinear} least squares. The corresponding {kernel} estimator is 
\begin{align}\label{eq:sst_est2}
\widehat{\Theta}(s) = \arg\min_{\Theta(s)}\ \sum_i\sum_t &\ \mathcal{K}_{{h}_2}(S_i,s)\Big( \widehat{y}_{it}^* - \beta_K(s) k_{it}-\beta_L(s)l_{it}\ - \notag \\
&\ \rho_0(s)- \rho_1(s) \Big[\widehat{\nu}^*_{it-1}-\beta_K(s) k_{it-1}-\beta_L(s)l_{it-1}\Big]+\rho_2(s)G_{it-1} \Big)^2,
\end{align}
where {$h_2$ is the number of nearest neighbors of a fixed location $s$ in the second-step estimation. It diverges faster than does the first-step smoothing parameter $h_1$} so that the first-step estimation has an asymptotically ignorable impact on the second step.

Lastly, the firm productivity is estimated as $\widehat{\omega}_{it}=y_{it}- \widehat{\beta}_K(S_i)k_{it}-\widehat{\beta}_L(S_i)l_{it}-\widehat{\beta}_M(S_i)m_{it}-\widehat{\eta}_{it}$ using the results from both steps.

\medskip

{
\noindent\textbf{Finite-Sample Performance.} Before applying our proposed methodology to the data, we first study its performance in a small set of Monte Carlo simulations. The results are encouraging, and simulation experiments show that our estimator recovers the true parameters well. As expected of a consistent estimator, the estimation becomes more stable as the sample size grows. For details, see Appendix \ref{sec:appex_sim}.

\medskip

\noindent\textbf{Inference.} Due to a multi-step nature of our estimator as well as the presence of nonparametric components, computation of the asymptotic variance of the estimators is not simple. For statistical inference, we therefore use bootstrap. We approximate sampling distributions of the estimators via wild residual block bootstrap that takes into account a panel structure of the data, with all the steps bootstrapped jointly owing to a sequential nature of our estimation procedure. The bootstrap algorithm is described in Appendix \ref{sec:appx_inference}. 

\medskip

\noindent\textbf{Testing of Location Invariance.} Given that our semiparametric locationally varying production model nests a more traditional fixed-parameter specification that implies locational invariance of the production function and the productivity evolution as a special case, we can formally discriminate between the two models to see if the data support our more flexible modeling approach. We discuss this specification test in detail in Appendix \ref{sec:appx_test}.
}

 
\section{Locational Productivity Differential Decomposition}
\label{sec:decomposition_level}

Since the production function can vary across space, a meaningful comparison of productivity for firms dispersed across space now requires that locational differences in technology be explicitly controlled. That is, the productivity differential between two firms is no longer limited to the difference in their firm-specific total factor productivities $\omega_{it}$ (unless they both belong to the same location) because either one of the firms may have access to a more productive technology $F_S(\cdot)$. Given that locational heterogeneity in production is the principal focus of our paper, in what follows, we provide a procedure for measuring and decomposing firm productivity differentials across any two locations of choice.

Let $\mathcal{L}(s,t)$ represent a set of $n_{t}^s$ firms operating in location $s$ in the year $t$. For each of these firms, the estimated Cobb-Douglas production function (net of random shocks) in logs is
\begin{equation}\label{eq:prodfn_hicks_cd-s}
	\widehat{y}_{it}^s = \widehat{\beta}_K(s) k_{it}^s+\widehat{\beta}_L(s)l_{it}^s+\widehat{\beta}_M(s)m_{it}^s + \widehat{\omega}_{it}^s, 
\end{equation}
where we have also explicitly indexed these firms' observable output/inputs as well as the estimated productivities using the location. Averaging over these firms, we arrive at the ``mean'' production function for location $s$ in time $t$:
\begin{equation}\label{eq:prodfn_hicks_cd-save}
\overline{y}_{t}^s = \widehat{\beta}_K(s) \overline{k}_{t}^s+\widehat{\beta}_L(s)\overline{l}_{t}^s+\widehat{\beta}_M(s)\overline{m}_{t}^s + \overline{\omega}_{t}^s, 
\end{equation}
where $\overline{y}_{t}^s= \tfrac{1}{n^s_t}\sum_i\widehat{y}_{it}^s\mathbbm{1}\{i\in \mathcal{L}(s,t)\}$, $\overline{x}_{t}^s=\tfrac{1}{n^s_t}\sum_i x_{it}^s\mathbbm{1}\{i\in \mathcal{L}(s,t)\}$ for $x\in \{k,l,m\}$, and $\overline{\omega}_{t}^s= \tfrac{1}{n^s_t}\sum_i\widehat{\omega}_{it}^s\mathbbm{1}\{i\in \mathcal{L}(s,t)\}$.

Taking the difference between \eqref{eq:prodfn_hicks_cd-save} and the analogous mean production function for the benchmark location of interest $\kappa$ in the same year, we obtain the mean \textit{output} differential between these two locations {(in logs)}:
\begin{equation}\label{eq:prodfn_hicks_cd-skdif}
\underbrace{\overline{y}_{t}^s - \overline{y}_{t}^{\kappa}}_{\Delta\overline{y}_{t}^{s,\kappa}}= \left[\widehat{\beta}_K(s) \overline{k}_{t}^s+\widehat{\beta}_L(s)\overline{l}_{t}^s +\widehat{\beta}_M(s)\overline{m}_{t}^s\right] -
\left[\widehat{\beta}_K(\kappa) \overline{k}_{t}^{\kappa}+\widehat{\beta}_L(\kappa)\overline{l}_{t}^{\kappa} +\widehat{\beta}_M(\kappa)\overline{m}_{t}^{\kappa}\right] 
+ \Big[ \overline{\omega}_{t}^s - \overline{\omega}_{t}^{\kappa}\Big] .
\end{equation}

To derive the mean \textit{productivity} differential (net of input differences) between these two locations, we add and subtract the $s$ location's production technology evaluated at the $\kappa$ location's inputs, i.e., \mbox{ $\left[\widehat{\beta}_K(s) \overline{k}_{t}^{\kappa}+\widehat{\beta}_L(s)\overline{l}_{t}^{\kappa} +\widehat{\beta}_M(s)\overline{m}_{t}^{\kappa}\right]$}, in \eqref{eq:prodfn_hicks_cd-skdif}:
\begin{align}\label{eq:prodfn_hicks_cd-decomp}
\Delta \overline{\text{PROD}}_{t}^{s,\kappa} &\equiv \Delta\overline{y}_{t}^{s,\kappa} - \widehat{\beta}_K(s) \Delta \overline{k}_{t}^{s,\kappa} -\widehat{\beta}_L(s)\Delta \overline{l}_{t}^{s,\kappa} -\widehat{\beta}_M(s)\Delta \overline{m}_{t}^{s,\kappa} \notag \\
& =\underbrace{\left[\widehat{\beta}_K(s)-\widehat{\beta}_K(\kappa)\right]  \overline{k}_{t}^{\kappa}+
\left[\widehat{\beta}_L(s)-\widehat{\beta}_L(\kappa)\right] \overline{l}_{t}^{\kappa} +
\left[\widehat{\beta}_M(s)-\widehat{\beta}_M(\kappa)\right] \overline{m}_{t}^{\kappa}}_{\Delta\overline{\text{TECH}}_{t}^{s,\kappa}} + \underbrace{\Big[ \overline{\omega}_{t}^s - \overline{\omega}_{t}^{\kappa}\Big]}_{\Delta\overline{\text{TFP}}_{t}^{s,\kappa}},
\end{align}
where  $\Delta\overline{x}_{t}^{s,\kappa}=\overline{x}_{t}^{s}-\overline{x}_{t}^{\kappa}$ for $x\in \{k,l,m\}$. 

Equation \eqref{eq:prodfn_hicks_cd-decomp} measures mean productivity differential across space and provides a \textit{counterfactual} decomposition thereof. By utilizing the counterfactual output that, given its location-specific technology, the average firm in location $s$ would have produced using the mean inputs employed by the firms in location $\kappa$ in year $t$ $\big[\widehat{\beta}_K(s) \overline{k}_{t}^{\kappa}+\widehat{\beta}_L(s)\overline{l}_{t}^{\kappa} +\widehat{\beta}_M(s)\overline{m}_{t}^{\kappa}\big]$, we are able to measure the locational differential in the mean productivity of firms in the locations $s$ and $\kappa$ that is \textit{un}explained by their different input usage: $\Delta \overline{\text{PROD}}_{t}^{s,\kappa}$. More importantly, we can then decompose this locational differential in the total productivity into the contribution attributable to the  difference in production technologies $\Delta\overline{\text{TECH}}_{t}^{s,\kappa}$ and to the  difference in the average total-factor {operations efficiencies} $\Delta\overline{\text{TFP}}_{t}^{s,\kappa}$.

The locational productivity differential decomposition in \eqref{eq:prodfn_hicks_cd-decomp} is time-varying, but should one be interested in a scalar measure of locational heterogeneity for the entire sample period, time-specific averages can be replaced with the  ``grand'' averages computed by pooling over all time periods.


\section{Empirical Application}
\label{sec:application}

Using our proposed model and estimation methodology, we explore the locationally heterogeneous production technology among manufacturers in the Chinese chemical industry. We report location-specific elasticity and productivity estimates for these firms and then decompose differences in their productivity across space to study if the latter are mainly driven by the use of different production technologies or the underlying total factor productivity differentials.


\subsection{Data}
\label{sec:data}

We use the data from \citet{baltagietal2016}. The dataset is a panel of $n=12,490$  manufacturers of chemicals continuously observed over the 2004--2006 period ($T=3$). The industry includes manufacturing of basic chemical materials (inorganic acids and bases, inorganic salts and organic raw chemical materials), fertilizers, pesticides, paints, coatings and adhesives, synthetic materials (plastic, synthetic resin and fiber) as well as daily chemical products (soap and cleaning compounds). The original source of these firm-level data is the Chinese Industrial Enterprises Database survey conducted by China's National Bureau of Statistics (NBS) which covers all state-owned firms and all non-state-owned firms with sales above 5 million Yuan (about \$0.6 million). \citet{baltagietal2016} have geocoded the location of each firm at the zipcode level in terms of the longitude and latitude (the ``$S$'' variables) using their postcode information in the dataset. The  coordinates are constructed for the location of each firm's headquarters and are time-invariant. By focusing on the continually operating firms, we mitigate a potential impact of spatial sorting (as well as the attrition due to non-survival) on the estimation results and treat the firm location as fixed (exogenous). The total number of observations is 37,470. 

Figure \ref{fig:number} shows the spatial distribution of firms in our dataset on a map of mainland China (we omit area in the West with no data in our sample). The majority are located on the East Coast and in the Southeast of China, especially around the Yangtze River Delta that is generally comprised of Shanghai and the surrounding areas, the southern Jiangsu province and the northern Zhejiang province. 

\begin{table}[t]
	\centering
	\caption{Data Summary Statistics}\label{tab:datasummary}
	\footnotesize
	\makebox[\linewidth]{
		\begin{tabular}{lrrrr}
			\toprule[1pt]
			Variables& Mean & 1st Qu. & Median  & 3rd Qu.\\
			\midrule
			&\multicolumn{4}{c}{\it \textemdash Production Function Variables\textemdash} \\[2pt]
			Output         & 86,381.98 & 11,021.09 & 23,489.53 & 59,483.71 \\
			Capital      & 35,882.40  & 1,951.47 & 5,319.28 & 17,431.35 \\
			Labor       & 199.07  & 43.00       & 80.00       & 178.00      \\
			Materials      & 48,487.82 & 5,896.49 & 12,798.35 & 33,063.81 \\[2pt]
			
			&\multicolumn{4}{c}{\it \textemdash Productivity Controls\textemdash} \\[2pt]
			Skilled Labor Share  	& 0.174    & 0.042    & 0.111    & 0.242    \\
			Foreign Equity Share 	& 0.140     & 0.000        & 0.000        & 0.000        \\
			Exporter        & 0.237    &   &   &      \\
			State-Owned   & 0.051    &   &   &       \\[2pt]
			
			&\multicolumn{4}{c}{\it \textemdash Location Variables\textemdash} \\[2pt]
			Longitude     & 2.041    & 1.984    & 2.068    & 2.102    \\
			Latitude    & 0.557    & 0.504    & 0.547    & 0.632    \\
			\midrule
			\multicolumn{5}{p{9.3cm}}{\scriptsize Output, capital and materials are in 1,000s of 1998 RMB. Labor is measured in the number of employees. The skilled labor share and foreign equity share are unit-free proportions. The exporter and state-owned variables are binary indicators. The location coordinates are in radians.} \\			
			\bottomrule[1pt]
		\end{tabular}
	}
\end{table}

The key production variables are defined as follows. Output ($Y$) is measured using sales. The labor input ($L$) is measured by the number of employees. Capital stock ($K$) is the net fixed assets for production and operation, and the materials ($M$) are defined as the expenditure on direct materials. Output, capital and materials are deflated to the 1998 values using the producer price index, the price index for investment in fixed assets and the purchasing price index for industrial inputs, respectively, where the price indices are obtained from the NBS. The unit of monetary values is thousands RMB (Chinese Yuan). 

We include four productivity-modifying variables in the evolution process of firm productivity $\omega_{it}$: the share of high-skilled workers ($G_1$), which is defined as the fraction of workers with a university or comparable education and is time-invariant because the data on workers' education level are only available for 2004; the foreign equity share ($G_2$), which is measured by the proportion of equity provided by foreign investors; a binary export status indicator ($G_3$), which takes value one if the firm is an exporter and zero otherwise; and a binary state/public ownership status indicator ($G_4$), which takes value one if the firm is state-owned and zero otherwise. 

Table \ref{tab:datasummary} shows the summary statistics, including the mean, 1st quartile, median and 3rd quartile for the variables. For the production-function variables, the mean values are significantly larger than their medians, which suggests that their distributions are skewed to the right. Among firms in the chemical industry, 23.7\% are exporters and 5.1\% are state-owned. Most firms do not have foreign investors, and the average ratio of foreign to total equity is 0.14. On average, 17.4\% of employees in the industry have a college degree or equivalent. 


\subsection{Estimation Results}
\label{sec:results}

In order to estimate the locationally varying (semiparametric) production function and firm productivity process in \eqref{eq:prodfn_hicks_cd}--\eqref{eq:productivity_hicks_lawsp}, we use the data-driven leave-one-location-out cross-validation method to choose the optimal {number of nearest neighboring locations in each step of the estimation ($h_1$ and $h_2$) to smooth over ``contextual'' location variables} $S_i$ inside the unknown functional coefficients. {This smoothing parameters regulate} spatial weighting of neighboring firms in kernel {fitting} and, as noted earlier, by selecting it via a data-driven procedure, we avoid the need to rely on \textit{ad hoc} specifications of both the spatial weights and radii defining the extent of {neighborhood influences}. The optimal {$h_1$ and $h_2$ values are 520 and 340 firm-years in the first- and second-step estimation, respectively. On average across all $s$, the corresponding adaptive bandwidths are 0.0171 and 0.0169 radians.} These bandwidth values are reasonable, given our sample size and the standard deviations of the longitude and latitude,\footnote{For the reference, the sample standard deviations for the longitude and latitude are respectively 0.0889 and 0.0941 radians.} and, evidently, are \textit{not} too large to ``smooth out'' the firm location {to imply location invariance/homogeneity. In fact, we can argue this more formally if kernel-smoothing is done using \textit{fixed} bandwidths so that we can rely on the theoretical results by \citet{halletal2007}, whereby local-constant kernel methods can remove irrelevant regressors via data-driven over-smoothing (i.e., by selecting large bandwidths). When we re-estimate our locationally-varying model in this manner, the optimal fixed bandwidths for the longitude and latitude in the first-step estimation are 0.009 and 0.010 radians, respectively; the corresponding second-step bandwidths are 0.024 and 0.023 radians. Just like in the case of adaptive bandwidths, these bandwidth values are fairly small relative to variation in the data, providing strong evidence in support of the overall relevancy of geographic location for firm production (i.e., against location invariance). Our location-varying formulation of the production technology and productivity is also formally supported by the \citet{ullah1985} specification test described in Appendix \ref{sec:appx_test}. Using cross-validated fixed bandwidths,} the bootstrap $p$-value is {0.001.} At the conventional significance level, our locationally heterogeneous production model is confidently preferred to a location-invariant formulation.

{In what follows, we discuss our semiparametric results obtained using adaptive bandwidths. For inference, we use the bias-corrected bootstrap percentile intervals as described in Appendix \ref{sec:appx_inference}. The number of bootstrap replications is set to $B=1,000$.}

\paragraph{Production Function.} We first report production-function estimates from our main model in which the production technology is locationally heterogeneous. We then compare these estimates with those obtained from the more conventional, {location-invariant} model that \textit{a priori} assumes common production technology for all firms. {The latter ``global'' formulation of the production function postulates constancy of the production relationship over space. This model is therefore fully parametric (with constant coefficients) and a special case of our locationally-varying model when $S_i$ is fixed across all $i$. Its estimation is straightforward and follows directly from \eqref{eq:fst_est2_pre}--\eqref{eq:sst_est} by letting the adaptive bandwidths in both steps diverge to $\infty$ which, in effect, obviates the need to locally weight the data because all kernels will be the same (for details, see Appendix \ref{sec:appx_test}).}\footnote{{Following a suggestion provided by a referee, we also estimate the location-invariant model {with location fixed effects added to the production function during the estimation.} We find the results do not change much from including these location effects, and therefore these results are not reported.}} 

\begin{table}[p]
	\centering
	\caption{Input Elasticity Estimates}\label{tab:coef}
	\footnotesize
	\makebox[\linewidth]{
		\begin{tabular}{lcccc|c}
			\toprule[1pt]
			& \multicolumn{4}{c}{\it Locationally Varying} & \it Location-Invariant \\
			& Mean & 1st Qu. & Median  & 3rd Qu.& Point Estimate\\
			\midrule	
			
			Capital                  & 0.112           & 0.095            & 0.115           & 0.128          & 0.130             \\
			& (0.104, 0.130)  & (0.083, 0.116)   & (0.110, 0.130)    & (0.119, 0.147) & (0.118, 0.141)   \\
			Labor                    & 0.303           & 0.272            & 0.293           & 0.342          & 0.299            \\
			& (0.285, 0.308)  & (0.248, 0.284)    & (0.278, 0.293)  & (0.313, 0.356) & (0.280, 0.318)   \\
			Materials                 & 0.480           & 0.452            & 0.481           & 0.503          & 0.495            \\
			& (0.466, 0.501)  & (0.414, 0.467)     & (0.437, 0.502)   & (0.456, 0.524) & (0.460, 0.519)   \\
			\midrule
			\multicolumn{6}{p{13cm}}{\scriptsize The left panel summarizes point estimates of $\beta_{\kappa}(S_i)\ \forall\ \kappa\in\{K, L, M\}$ with the corresponding two-sided 95\% bias-corrected confidence intervals in parentheses. The right panel reports their counterparts from a fixed-coefficient location-invariant model. } \\
			\bottomrule[1pt]
		\end{tabular}
	}
\end{table}

\begin{figure}[p]
	\centering
	\includegraphics[scale=0.36]{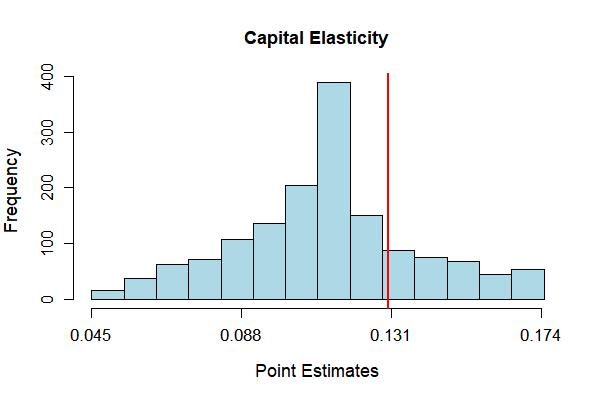}\includegraphics[scale=0.36]{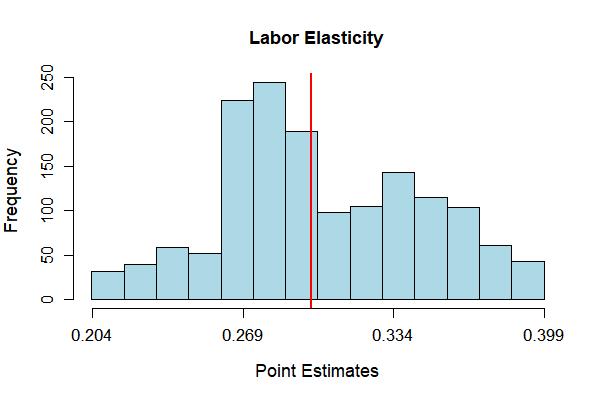}
	\includegraphics[scale=0.36]{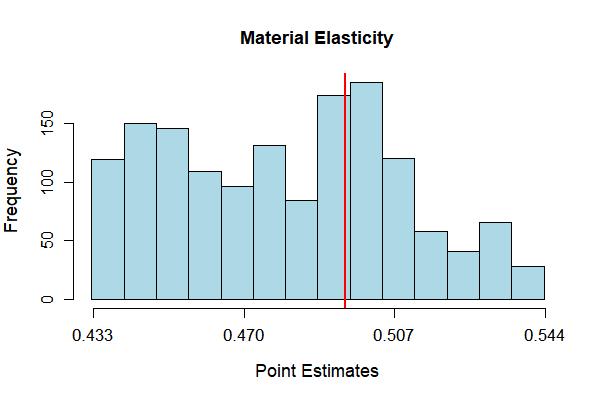}\includegraphics[scale=0.36]{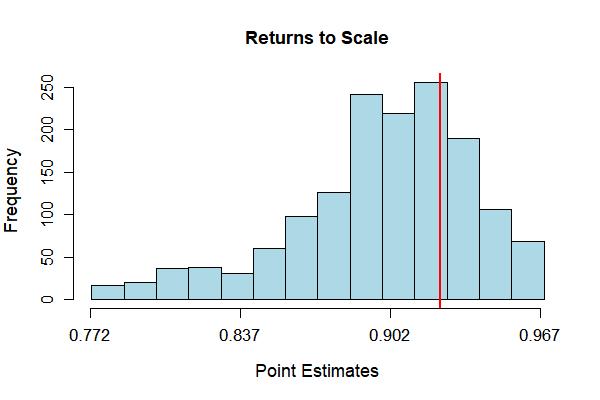}
	\caption{Input Elasticity Estimates \\ {\small (Notes: Vertical lines correspond to location-invariant estimates)}} \label{fig:coef_elas}
\end{figure}

Since our model has location-specific input elasticities, there is a distribution of them (over space) and Table \ref{tab:coef} summarizes their point estimates. The table also reports the elasticity estimates from the alternative, location-invariant model. The corresponding two-sided 95\% bias-corrected confidence intervals for these statistics are reported in parentheses. Based on our model, the mean (median) capital, labor and material elasticity estimates are {0.112, 0.303 and 0.480 (0.115, 0.293 and 0.481),} respectively. Importantly, these location-specific elasticities show significant variation. {For the capital and labor inputs, the first quartiles are significantly different from the third quartiles.} Within the inter-quartile interval of their point estimates, elasticities of capital, labor and materials respectively increase by {0.033, 0.070 and 0.051, which in turn correspond to the 35\%, 26\% and 11\% changes.} 

In comparison, the elasticity estimates from the location-invariant production function with fixed coefficients are all larger than the corresponding {median} estimates from our model and fall in between the second and third quartiles of our locationally-varying point estimates. Figure \ref{fig:coef_elas} provides visualization of the non-negligible technological heterogeneity in the chemicals production technology across different locations in China, which the traditional location-invariant model assumes away. The figure plots histograms of the estimated location-specific input elasticities (and the returns to scale) with the location-invariant counterpart estimates depicted by vertical lines. Consistent with the results in Table \ref{tab:coef}, all distributions show relatively wide dispersion, and the locationally homogeneous {model} is apparently unable to provide a reasonable representation of production technology across different regions.

\begin{table}[t]
	\centering
	\caption{Locationally Varying Returns to Scale Estimates}\label{tab:RTC}
	\footnotesize
	\makebox[\linewidth]{
		\begin{tabular}{lcccc|cc}
			\toprule[1pt]
			& Mean & 1st Qu. & Median  & 3rd Qu.& $= 1$ &$<1$ \\
			\midrule	
			RTS & 0.895          & 0.875          & 0.903         & 0.929              & 21.6\%            &  82.3\%        \\
			& (0.820, 0.931) & (0.801, 0.912) & (0.827, 0.942) & (0.855, 0.968) & &  \\
			\midrule
			\multicolumn{7}{p{11.8cm}}{\scriptsize The left panel summarizes point estimates of $\sum_{\kappa}\beta_{\kappa}(S_i)$ with $\kappa\in\{K, L, M\}$ with the corresponding two-sided 95\% bias-corrected confidence intervals in parentheses. The counterpart estimate of the returns to scale from a fixed-coefficient location-invariant model is 0.924 (0.865, 0.969). The right panel reports the shares of locations in which location-specific point estimates are (\textit{i}) not significantly different from 1 (constant returns to scale) and (\textit{ii}) statistically less than 1 (decreasing returns to scale). The former classification is based on a two-sided test, the latter is on a one-sided test.} \\
			\bottomrule[1pt]
		\end{tabular}
	}
\end{table}

Table \ref{tab:RTC} provides summary statistics of the estimated returns to scale (RTS) from our locationally varying production function (also see the bottom-right plot in Figure \ref{fig:coef_elas}). The mean RTS is {0.895, and the median is 0.903, with the inter-quartile range being 0.054.} The right panel of Table \ref{tab:RTC} reports the fraction of locations in which the Chinese manufacturers of chemicals exhibit constant or decreasing returns to scale. This classification is based on the RTS point estimate being statistically equal to or less than one, respectively, at the 5\% significance level. The ``$=1$'' classification is based on a two-sided test, whereas the ``$<1$'' test is one-sided. In most locations in China {(82.3\%)}, the production technologies of the chemicals firms exhibit \textit{dis}economies of scale, but {21.6\%} regions show evidence of the constant returns to scale (i.e., scale efficiency).   

\begin{figure}[t]
	\centering
	\includegraphics[scale=0.3]{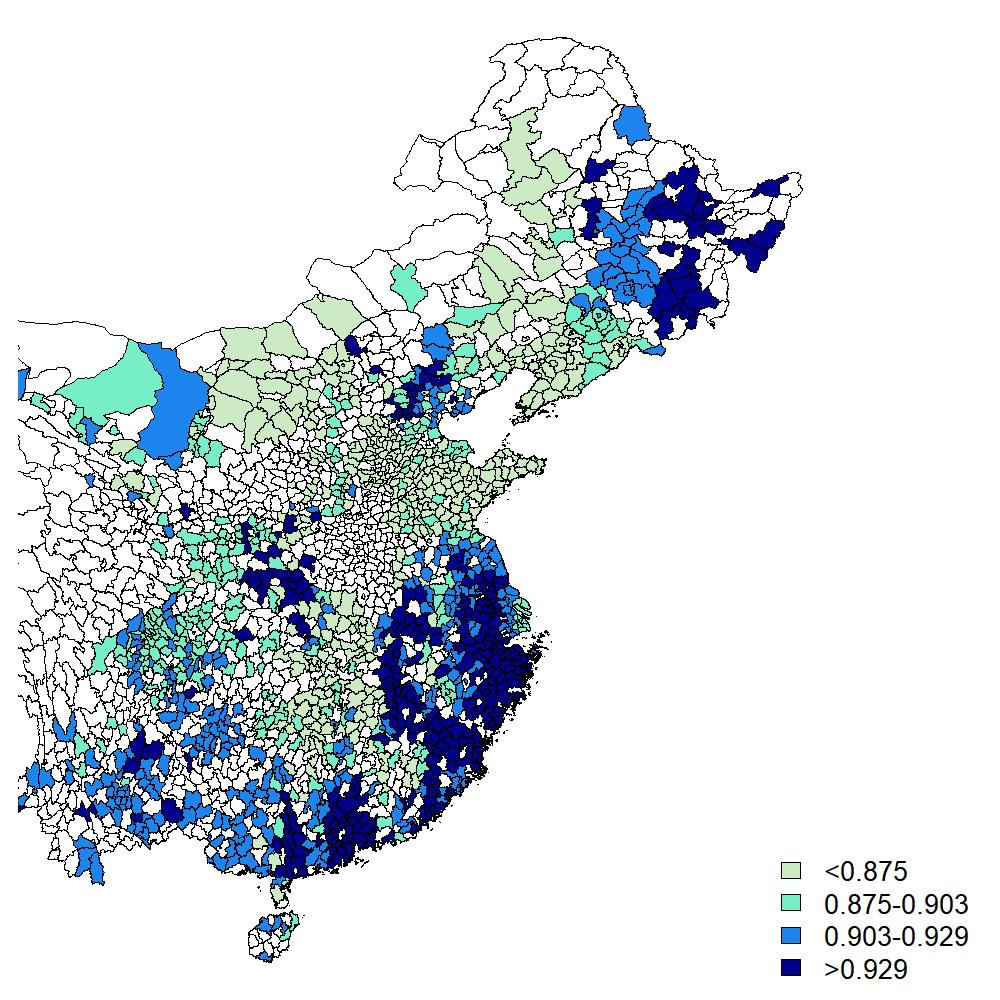}
	\caption{Spatial Distribution of Returns to Scale Estimates \\
		{\small	(Notes: The color shade cutoffs correspond to the first, second (median) and third quartiles)}  } \label{fig:rts}
\end{figure}

To further explore the locational heterogeneity in the production technology for chemicals in China, we plot the spatial distribution of the RTS estimates in the country in Figure \ref{fig:rts}. We find that the firms {with the largest RTS are mainly located in the Southeast Coast provinces and some parts of the West and Northeast China.} The area nearby Beijing also exhibits larger RTS. There are a few possible explanations of such a geographic distribution of the returns to scale. As noted earlier, spillovers and agglomeration have positive effects on the marginal productivity of inputs which typically take form of the scale effects, and they may explain the high RTS on the Southeast Coast and in the Beijing area. Locality-specific resources, culture and polices can also facilitate firms' production process. For example, the rich endowment of the raw materials like coal, phosphate rock and sulfur make the provinces such as Guizhou, Yunnan and Qinghai among the largest fertilizer production zones in China. Furthermore, RTS is also related to the life cycle of a firm. Usually, it is the small, young and fast-growth firms that enjoy higher RTS, whereas the more mature firms that have grown bigger will have transitioned to the low-RTS scale. This may explain the prevalence of the higher-RTS firms in the West and Northeast China.

\paragraph{Productivity Process.} We now analyze our semiparametric estimates of the firm productivity process in \eqref{eq:productivity_hicks_lawsp}. Table \ref{tab:prod.coef} summarizes point estimates of  the location-specific marginal effects of productivity determinants in the evolution process of $\omega_{it}$, with the corresponding two-sided 95\% bias-corrected confidence intervals in parentheses. In the last column of the left panel, for each productivity-enhancing control $G_{it}$, we also report the share of locations in which location-specific point estimates are statistically positive (at a 5\% significance level) as inferred via a one-sided test.

\begin{table}[p]
	\centering
	\caption{Productivity Process Coefficient Estimates}\label{tab:prod.coef}
	\footnotesize
	\makebox[\linewidth]{
		\begin{tabular}{lccccc|c}
			\toprule[1pt]
			          & \multicolumn{5}{c}{\it Locationally Varying} & \it Location-Invariant \\
			Variables& Mean & 1st Qu. & Median  & 3rd Qu.& $>0$ & Point Estimate\\
			\midrule	
			Lagged Productivity        & 0.576 & 0.518 & 0.597 & 0.641 & 99.9\% & 0.497          \\
			& (0.540, 0.591) & (0.469, 0.541) & (0.553, 0.614) & (0.580, 0.665) &   & (0.455, 0.530)   \\
			Skilled Labor Share & 0.387 & 0.287 & 0.419 & 0.500  & 85.7\%  & 0.387            \\
			& (0.346, 0.395) & (0.241, 0.309) & (0.345, 0.459) & (0.471, 0.493) &   & (0.345, 0.425)    \\
			Foreign Equity Share         & 0.054 & --0.001 & 0.062 & 0.103 & 47.7\% & 0.056		            \\
			& (0.006, 0.074) & (--0.034, 0.066) & (0.033, 0.069) & (0.099, 0.099) &  & (0.036, 0.075) 	   \\
			Exporter              & --0.001 & --0.032 & --0.005 & 0.038 & 24.0\% & 0.006 	            \\
			& (--0.011, 0.018) & (--0.041, --0.016) & (--0.012, 0.013) & 	(0.025, 0.067) &  & (--0.008, 0.018) 	  \\
			State-Owned           & 0.005 & --0.052 & 0.007 & 	0.073 & 	29.6\% & 	--0.043            \\
			& (--0.021, 0.010) & 	(--0.101, --0.014) & (--0.028, 0.025) & 	(0.062, 0.076) &  & (--0.072, --0.009) 		 \\
			\midrule
			\multicolumn{7}{p{16.5cm}}{\scriptsize The left panel summarizes point estimates of $\rho_j(S_i)\ \forall\ j=1,\dots,\dim(G)$ with the corresponding two-sided 95\% bias-corrected confidence intervals in parentheses. Reported is also a share of locations in which location-specific point estimates are statistically positive as inferred via a one-sided test. The right panel reports the counterparts from a fixed-coefficient location-invariant model.} \\
			\bottomrule[1pt]
		\end{tabular}
	}
\end{table}

\begin{figure}[p]
	\centering
	\includegraphics[scale=0.33]{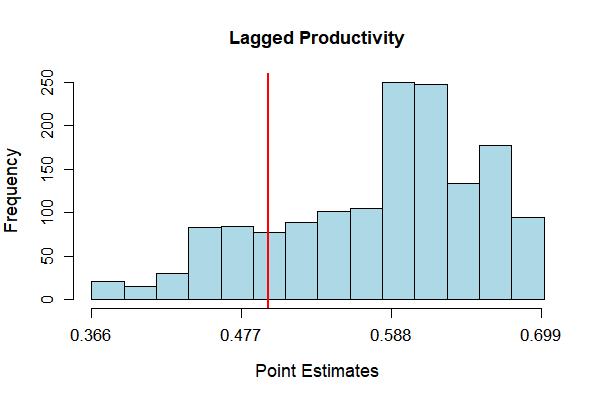}
	\includegraphics[scale=0.33]{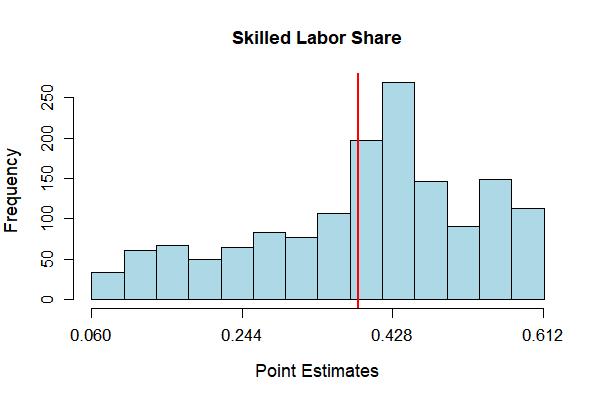}\includegraphics[scale=0.33]{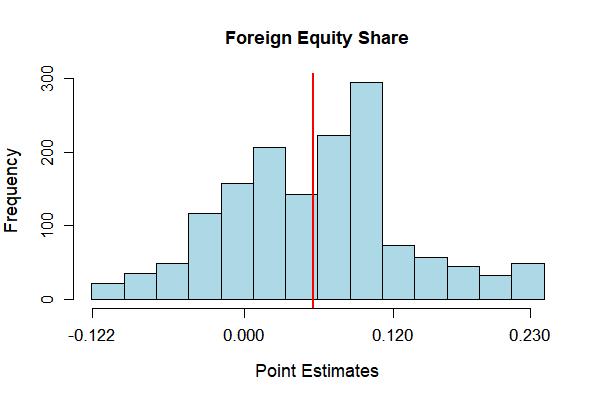}
	\includegraphics[scale=0.33]{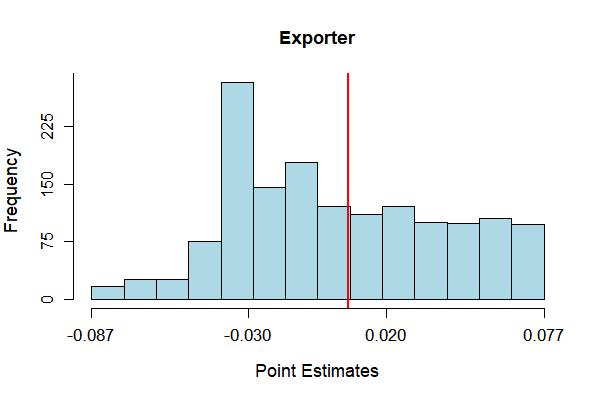}\includegraphics[scale=0.33]{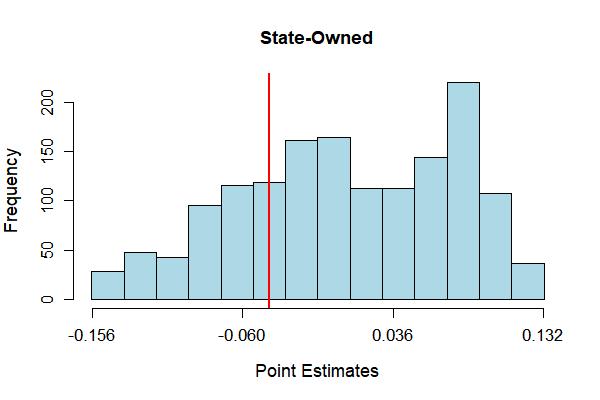}	
	\caption{Productivity Process Coefficient Estimates \\ {\small (Notes: Vertical lines correspond to location-invariant estimates)}} \label{fig:coef_prod}
\end{figure}

The autoregressive coefficient on the lagged productivity, which measures the persistence of $\omega_{it}$, is {0.576 at the mean and 0.597 at the median, with the quartile statistics varying from 0.518 to 0.641.} It is significantly positive for firms in {virtually} all locations. For firms in most locations {(85.7\%)}, skilled labor has a large and significantly positive effect on productivity: a percentage point increase in the skilled labor share is associated with an improvement in the next period's firm productivity by about 0.4\%, on average. Point estimates of the foreign ownership effect are positive {in the majority of locations}, but firms in only about half the locations benefit from a statistically positive productivity-boosting effect of the inbound foreign direct investment, with the average magnitude of only {7/50} of that attributable to hiring more skilled labor. In line with the empirical evidence reported for China's manufacturing in the literature \citep[see][and references therein]{malikovetal2020}, firms in most regions show insignificant {(and negative)} effects of the export status on productivity. The ``learning by exporting'' effects are very limited and statistically positive in {a quarter} of locations only. Interestingly, we find that state/public ownership is a significantly positive contributor to the improvements in firm productivity in about a third of the locations in which the Chinese chemicals manufacturing firms operate. This may be because the less productive state firms exited the market during the market-oriented transition in the late 1990s and early 2000s (prior to our sample period), and the remaining state-owned firms are larger and more productive \citep[also see][]{hsiehsong2015,zhaoqiankumbhakar2020}. Another potential reason may be that state ownership could have brought non-trivial financing benefits to these firms which, otherwise, were usually financially constrained due to the under-developed financial market in China during that period. 

The far right panel of Table \ref{tab:prod.coef} reports productivity effects of the $G_{it}$ controls estimated using the location-invariant model. Note that, under the assumption of a location-invariant production, the evolution process of $\omega_{it}$ becomes a parametric linear model, and there is only one point estimate of each fixed marginal effect for all firms. Comparing these estimates with the {median} estimates from our model, the location-invariant marginal effects tend to be smaller. While the persistence coefficient as well as fixed coefficients on the skilled labor and foreign equity shares are positive and statistically significant, the location-invariant estimate of the state ownership effect on productivity is however significantly negative (for all firms, by design). Together with the tendency of a location-invariant model to underestimate, this underscores the importance of allowing sufficient flexibility in modeling heterogeneity across firms (across different locations, in our case) besides the usual Hicks-neutral TFP. 

The contrast between the two models is even more apparent in Figure \ref{fig:coef_prod}, which plots the distributions of estimated marginal effects of the productivity-enhancing controls. Like before, the location-invariant counterparts are depicted by vertical lines. The distribution of each productivity modifier spans a relatively wide range, and the corresponding location-invariant estimates are evidently not good representatives for the {centrality} of these distributions. For example, the productivity-boosting effect of the firm's skilled labor roughly varies between {0.06 and 0.61\% per unit percentage point increase in the skilled labor share, depending on the location.} The distribution of this marginal effect across locations is somewhat left-skewed, and the corresponding location-invariant effect estimate evidently does not measure central tendency of these locationally-varying effects well. Similar observations can be made about other varying coefficients in the productivity process.

\paragraph{Productivity Decomposition.} We now examine the average productivity differentials for firms in different regions. To this end, we perform the locational decomposition proposed in Section \ref{sec:decomposition_level} to identify the sources of production differences that cannot be explained by input usage. Recall that, by our decomposition, the locational differential in the mean total productivity ($\Delta \overline{\text{PROD}}^{s,\kappa}_t$) accounts for the cross-regional variation in both the input elasticities ($\Delta\overline{\text{TECH}}_{t}^{s,\kappa}$) and the total factor productivity ($\Delta\overline{\text{TFP}}^{s,\kappa}$). It is therefore more inclusive than the conventional analyses that rely on fitting a common production technology for all firms regardless of their locations and thus confine cross-firm heterogeneity to differences in $\omega_{it}$ only.

\begin{table}[t]
	\centering
	\caption{Locational Productivity Differential Decomposition}\label{tab:decom}
	\footnotesize
	\makebox[\linewidth]{
		\begin{tabular}{lcccc}
			\toprule[1pt]
			Components& Mean & 1st Qu. & Median  & 3rd Qu.\\
			\midrule	
			& \multicolumn{4}{c}{\it Locationally Varying Model} \\
			$\Delta\overline{\text{TECH}}^{s,\kappa}$                 & 1.292 & 1.135 & 1.331 & 1.521         \\
			$\Delta\overline{\text{TFP}}^{s,\kappa}$                   & 0.574 & 0.203 & 0.571 & 0.893        \\
			$\Delta \overline{\text{PROD}}^{s,\kappa}$              & 1.866 & 1.652 & 1.869 & 2.086         \\
			\midrule
			& \multicolumn{4}{c}{\it Location-Invariant Model} \\
			$\Delta \overline{\text{PROD}}^{s,\kappa}$ & 1.797 & 1.589 & 1.816 & 2.040          \\
			\midrule
			\multicolumn{5}{p{7cm}}{\scriptsize The top panel summarizes point estimates of the locational mean productivity differential $\Delta \overline{\text{PROD}}^{s,\kappa}=\Delta\overline{\text{TECH}}^{s,\kappa}+ \Delta\overline{\text{TFP}}^{s,\kappa}$ with the corresponding two-sided 95\% bias-corrected confidence intervals in parentheses. The bottom panel reports the counterparts from a fixed-coefficient location-invariant model for which, by construction, $\Delta \overline{\text{PROD}}^{s,\kappa}= \Delta\overline{\text{TFP}}^{s,\kappa}$ with $\Delta\overline{\text{TECH}}^{s,\kappa}=0$. In both cases, the decomposition is pooled for the entire sample period and the benchmark/reference location $\kappa$ is the one with the smallest mean production: $\kappa=\arg\min_{s} \overline{y}^{s}$.} \\
			\bottomrule[1pt]
		\end{tabular}
	}
\end{table}

Table \ref{tab:decom} presents the decomposition results (across locations $s$) following \eqref{eq:prodfn_hicks_cd-decomp}. Because we just have three years of data, we perform the decomposition by pooling over the entire sample period. Thus, reported are the average decomposition results across 2002--2004. Also note that, for a fixed benchmark location $\kappa$, the decomposition is done for each $s$-location separately. For the benchmark/reference location $\kappa$, we choose the zipcode with the smallest mean production, i.e., $\kappa=\arg\min_{s} \overline{y}^{s}$, where $\overline{y}^{s}$ is defined as the time average of  \eqref{eq:prodfn_hicks_cd-save}.\footnote{Obviously, the choice of a reference location is inconsequential because its role is effectively that of a normalization.} Therefore, the numbers ($\times 100\%$) in Table \ref{tab:decom} can be interpreted as the percentage differences between the chemicals manufacturers operating in various locations ($s$) versus those from the least-production-scale region $(\kappa)$ in China. Because the reference location is fixed, the results are comparable across $s$.

Based on our estimates, the mean productivity differential is {1.866,} which means that, compared to the location with the smallest scale of chemicals production, other locations are, on average, {187\%} more productive (or more effective in the input usage). {The inter-quartile range of the average productivity differential spans from 1.652 and 2.086. Economically, these differences are large: firms that are located at the third quartile of the locational productivity distribution are about 43\% more productive than firms at the first quartile.} When we decompose the productivity differential into the technology and TFP differentials, on average, {$\Delta\overline{\text{TECH}}^{s,\kappa}$ is 2.3 times as large as $\Delta\overline{\text{TFP}}^{s,\kappa}$ and accounts for about 69\%} of the total productivity differences across locations.\footnote{That is, the ratio of $\Delta\overline{\text{TECH}}^{s,\kappa}$ to $\Delta\overline{\text{PROD}}^{s,\kappa}$ is 0.69.} This suggests that the cross-location {technological} heterogeneity in China's chemicals industry explains most of the productivity differential and that the regional {TFP differences are \textit{relatively} more modest.}

Table \ref{tab:decom} also summarizes the locational productivity differential estimates from the standard location-invariant model. Given that this model assumes fixed coefficients (same technology for all firms), we cannot perform a decomposition here, and all cross-location variation in productivity is \textit{a priori} attributed to TFP by design. Compared with our locationally-varying model, this model {yields similar} total productivity differentials across regions but, due to its inability to recognize technological differences, it {grossly} over-estimates cross-location differences in TFP.  

\begin{figure}[t]
	\centering
	\includegraphics[scale=0.29]{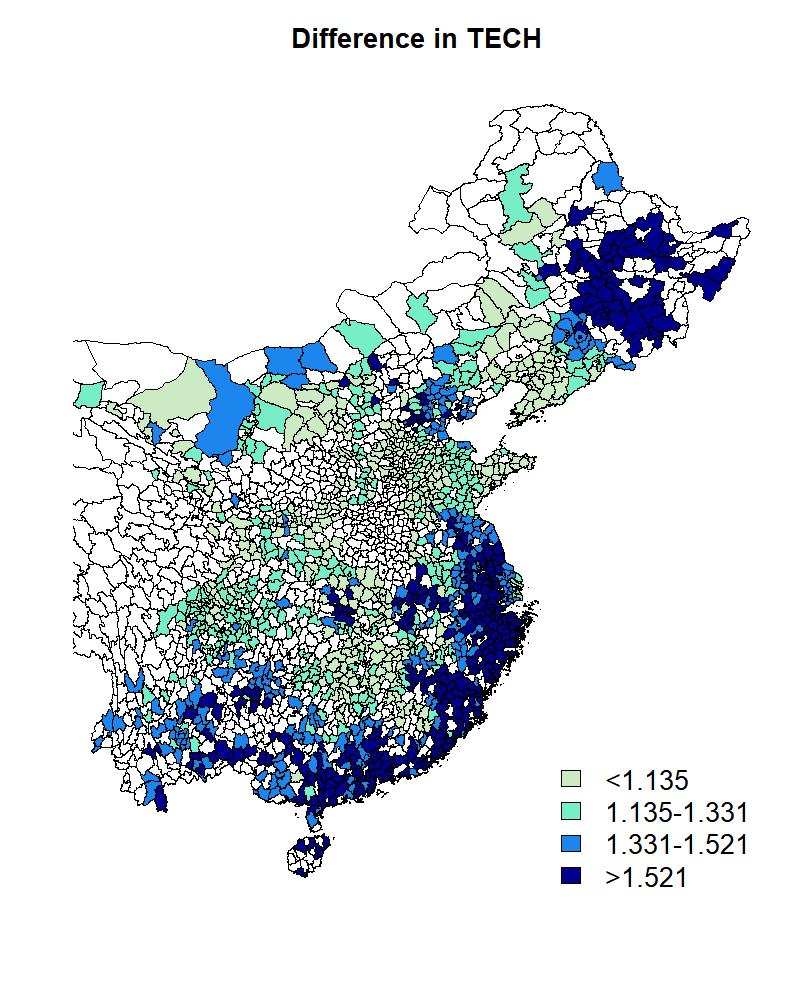}\includegraphics[scale=0.29]{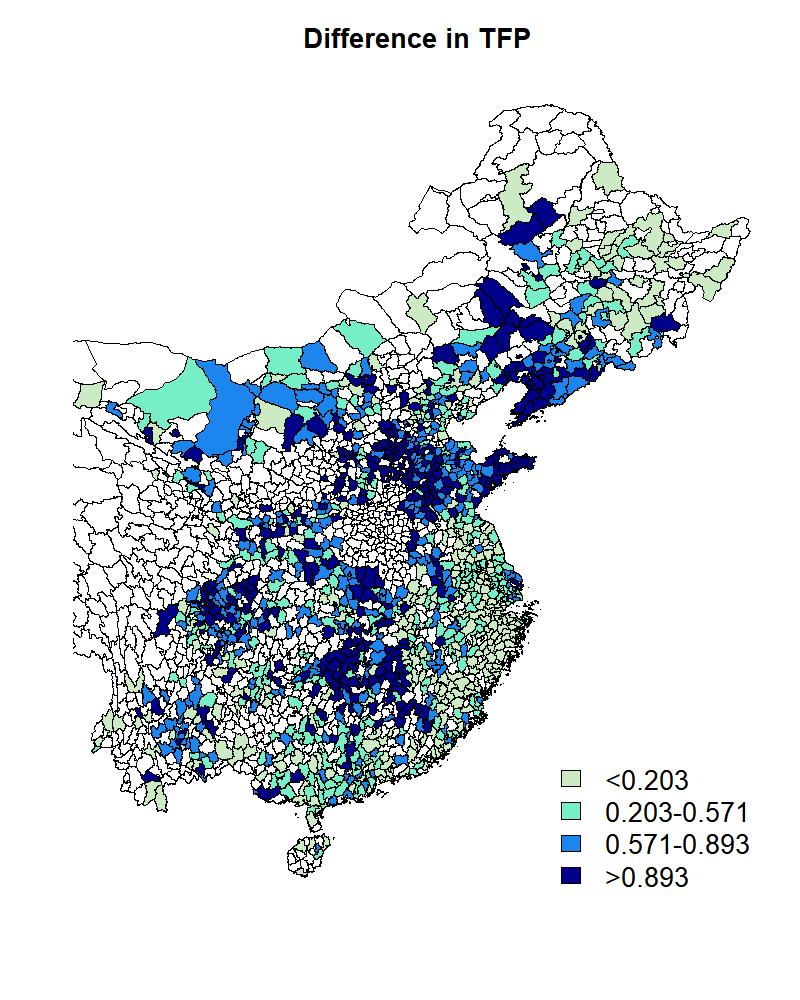}
	\caption{Locational Productivity Differential Decomposition Estimates Across Space \\ {\small (Notes: The color shade cutoffs correspond to the first, second (median) and third quartiles)} } \label{fig:decomposition_map}
\end{figure}

To explore the spatial heterogeneity in the decomposition components, we plot the spatial distributions of $\Delta\overline{\text{TECH}}^{s,\kappa}$, $\Delta\overline{\text{TFP}}^{s,\kappa}$ and $\Delta \overline{\text{PROD}}^{s,\kappa}$ on the map in Figure \ref{fig:decomposition_map}. The spatial distribution of $\Delta\overline{\text{TECH}}^{s,\kappa}$ aligns {remarkably} with that of RTS in Figure \ref{fig:rts}. Noticeably, the regions of agglomeration in the chemicals industry (see Figure \ref{fig:number}) tend to demonstrate large technology differentials. In contrast, the spatial distribution of $\Delta\overline{\text{TFP}}^{s,\kappa}$ shows quite a different pattern, whereby the locations of large TFP (differentials) are less concentrated. Unlike with the $\Delta\overline{\text{TECH}}^{s,\kappa}$ map, the dark-shaded regions on the $\Delta\overline{\text{TFP}}^{s,\kappa}$ map are widely spread around and have no clear overlapping with the main agglomeration regions in the industry. The comparison between these two maps suggests that, at least for the Chinese chemicals manufacturing firms, the widely-documented agglomeration effects on firm productivity are associated more with the scale effects via production technology rather than the improvements in overall TFP. That is, by locating closer to other firms in the same industry, it may be easier for a firm to pick up production technologies and know-hows that improve productiveness of inputs technologically and thus expand the input requirement set corresponding to the firm's output level\footnote{And more generally, shifting the \textit{family} of firm's isoquants corresponding to a fixed level of $\omega_{it}$ toward the origin.} \textit{given} its total factor productivity. Instead, agglomeration effects that increase the effectiveness of transforming all factors into the outputs via available technology (by adopting better business practices or management strategies) may be less likely to spill among the Chinese manufacturers of chemicals. Importantly, if we \textit{a priori} assume the fixed-coefficient production function common to all firms, the technological effects of agglomeration (via input elasticities) would be wrongly attributed to the TFP differentials. 


\section{Concluding Remarks}
\label{sec:conclusion}

Although it is widely documented {in the operations management literature} that the firm's location matters for its performance, few empirical studies {of operations efficiency} explicitly control for it. {This paper fills in this gap by providing a semiparametric methodology for the} identification of production functions in which locational factors have heterogeneous effects on the firm's production technology and productivity evolution. {Our approach is novel in that we
explicitly model spatial variation in parameters in the production-function estimation.} We generalize the popular Cobb-Douglas production function in a semiparametric fashion by writing the input elasticities and productivity parameters as unknown functions of the firm's {geographic} location. In doing so, not only do we render the production {technology} location-specific but also accommodate {neighborhood influences on firm operations} with the strength thereof depending on the distance between {firms. Importantly, this enables us to examine the role of cross-location differences in explaining the variation in operational productivity among firms. 

The proposed model is superior to the alternative SAR-type production-function formulations because it (i) explicitly estimates the locational variation in production functions, (ii) is readily reconcilable with the conventional production axioms and, more importantly, (iii) can be identified from the data by building on the popular proxy-variable methods, which we extend to incorporate locational heterogeneity in firm production.} Our methodology provides a practical tool for examining the effects of agglomeration and technology spillovers on firm performance {and will be most useful for empiricists focused on the analysis of operations efficiency/productivity and its ``determinants.''}

Using the methods proposed in our paper, we can {separate the effects of firm location on} production technology from those on firm productivity and find evidence consistent with the conclusion that agglomeration economies affect {the productivity of Chinese chemicals manufacturers mainly through the scale effects of production \textit{technology} rather than the improvements in overall TFP.} Comparing our flexible semiparametric model with the more conventional parametric model that postulates a common technology for all firms regardless of their location, we show that the latter does not provide {an adequate representation of} the industry and that the conclusion based on its results can be misleading. {For managerial implications, our study re-emphasizes the importance of firm location for its operations efficiency in manufacturing industries. Our findings also suggest that hiring skilled labor has a larger productivity effect compared to other widely-discussed productivity-enhancing techniques, such as learning by exporting. }


\appendix
\section*{Appendix}

\renewcommand\thetable{\thesection.\arabic{table}}
\renewcommand\thefigure{\thesection.\arabic{figure}}

\section{The SAR Production-Function Models}
\label{sec:appx_sar}

{A popular approach to incorporating locational/spatial effects in production models relies on spatial econometric techniques, whereby spatially-weighted averages of other firms' outputs (and sometimes inputs too) are included as additional regressors in the SAR production-function models. Not only does such a SAR specification of the production relationship continues to implausibly assume the common technology for all locations, it also becomes problematic in practice, when it comes to its estimation: (\textsl{i}) the SAR production functions imply additional, highly nonlinear parameter restrictions necessary to ensure that the conventional axioms of the production theory are not violated, albeit these are usually ignored in applied work, and (\textsl{ii}) the identification of SAR production-function models from data is hardly guaranteed due to the inapplicability of available proxy-variable estimators and the general lack of valid external instruments. In what follows, we discuss each of these considerations in detail.

\medskip

\noindent\textbf{Axiomatic Considerations.}\textemdash For expositional simplicity, for now let us assume a deterministic log-linear (Cobb-Douglas) input-output relationship with a single input and suppress the time index. The SAR production function a l\`a \citet{glassetal2016} in logs is given by
\begin{equation}\label{eq:sar1}
	y_i = \rho \sum_{j(\ne i)=1}^n d_{ij} y_j + \beta_K k_i + \omega_i,
\end{equation}
where $\{d_{ij}\ge0\}$ are the non-negative spatial weights that describe the architecture of inter-firm spatial dependence. Following the convention, $d_{ii}=0\ \forall i$ and $d_{ij}=d_{ji}\ \forall i, j$.  Putting these weights for all firms together gives a symmetric $n\times n$ non-stochastic spatial weighting matrix $\mathbf{D}$ which, following the popular practice, is row-standardized.\footnote{That is, $\sum_{j=1}^nd_{ij}=1$ for all rows $i=1,\dots,n$, although there are alternative normalizations available.} Assume that $\mathbf{D}$ is known. 
To ensure spatial stationarity, the spatial lag parameter is $\rho\in(-1,1)$.

The above formulation implies that the $i$th firm's output $y_i$ is a function of not only its own input $k_i$ but of all its neighbors' inputs $\{k_{j}\}$, and the elasticity of own capital is no longer equal to $\beta_K$. The latter are not without implications for the theoretical regularity conditions routinely assumed\textemdash explicitly or implicitly\textemdash about the production frontier. 

To make matters more concrete, letting $\boldsymbol{y}=(y_1,\dots,y_n)'$, $\boldsymbol{k}=(k_1,\dots,k_n)'$ and $\boldsymbol{\omega}=(\omega_1,\dots,\omega_n)'$, which all are the $n\times 1$ vectors, we have a vector of reduced-form production functions for all firms:
\begin{equation}
 	\boldsymbol{y} = \overbrace{\left[ \mathbf{I}_n-\rho \mathbf{D}\right]^{-1}}^{\mathbf{Q}} \left( \beta_K \boldsymbol{k} + \boldsymbol{\omega}\right),
\end{equation}
from where we have that, for each firm $i$, the production function is 
\begin{equation}\label{eq:rflog}
	y_i = \beta_K\sum_{j=1}^n Q_{ij} k_j + \sum_{j=1}^n Q_{ij}\omega_j,
\end{equation}
with $Q_{ij}$ being the $(i,j)$th element of the $n\times n$ ``spatial multiplier matrix'' $\mathbf{Q}$. Exponentiating \eqref{eq:rflog}, we arrive at the firm $i$'s production function in levels:
\begin{equation}\label{eq:rflevel}
	Y_i = K_i^{\beta_KQ_{ii}}\left[\prod_{j(\ne i)=1}^n K_j^{Q_{ij}} \right]^{\beta_K}  \prod_{j=1}^n \exp\{Q_{ij}\omega_j\}.
\end{equation}

Conventional production axioms imply, among other theoretical regularity conditions, the following about the shape of production function: monotonicity and concavity in inputs. Now, because the theory of production does not normally consider the possibility of inter-firm spillovers, technically it does not differentiate between the production unit's \textit{own} and its \textit{neighbors}' inputs. By implication however, the standard regularity conditions describe the within-unit technological relationships between the production unit's \textit{own} inputs and outputs. 

It is straightforward to see that \eqref{eq:rflevel} remains monotone in $K_i$ so long as the capital elasticity parameter is positive ($\beta_K>0$):
\begin{equation}\label{eq:rflevel_mono}
	\frac{\partial Y_i}{\partial K_i} = \beta_KQ_{ii}K_i^{\beta_KQ_{ii}-1}\left[\prod_{j(\ne i)=1}^n K_j^{Q_{ij}} \right]^{\beta_K}  \prod_{j=1}^n \exp\{Q_{ij}\omega_j\} >0
\end{equation}
because $K_i\in \Re_{+}$ and $Q_{ii}\in \Re_{+}$ for all $i$.

Things are not as trivial when it comes to curvature. Consider the own second partial derivative w.r.t.~capital:
\begin{equation}\label{eq:rflevel_concave}
 	\frac{\partial^2 Y_i}{\partial K_i^2} = \beta_KQ_{ii}\left(\beta_KQ_{ii}-1\right)K_i^{\beta_KQ_{ii}-2}\left[\prod_{j(\ne i)=1}^n K_j^{Q_{ij}} \right]^{\beta_K}  \prod_{j=1}^n \exp\{Q_{ij}\omega_j\}.
\end{equation}

For \eqref{eq:rflevel_concave} to remain non-positive in order to guarantee concavity of the production frontier, it needs be that the diagonal elements of the spatial multiplier matrix $\mathbf{Q}=\left[\mathbf{I}_n-\rho \mathbf{D}\right]^{-1}$ are such that $Q_{ii} \le \beta_K^{-1}$ (so long as $\beta_K>0$ just like earlier). Since $Q_{ii}$ is a function of $\rho$ and $\mathbf{D}$, the latter places restrictions on both the strength of spatial dependence\textemdash the magnitude of $\rho$\textemdash and the architecture of spatial relationships, i.e., the specification/design of $\mathbf{D}$. 

This is an important implication because, depending on the value of capital elasticity $\beta_K$, these restrictions can be quite strict if the acceptable range of $Q_{ii}$ is especially narrow. In fact, the permissible range is not $Q_{ii} \le \beta_K^{-1}$ but 
\begin{equation}\label{eq:Qconstr}
    1 \le Q_{ii} \le \beta_K^{-1}\quad \forall\ i,
\end{equation}
since the diagonal elements of $\mathbf{Q}$ are all no smaller than 1 by construction \citep[see][]{elhorst2010}. This already rules out the possibility of constant (internal) returns to scale commonly used in the applied productivity research, particularly at the aggregate level. Namely, if $\beta_K=1$, it must be that $Q_{ii}=1$ for all $i$ which is possible only if $\rho=0$. Consequently, cross-firm spatial spillovers in \eqref{eq:sar1} are consistent only with the decreasing returns to scale, i.e., when $\beta_K<1$. 

The practical implementation can get complicated even when one is willing to assume the decreasing returns to scale \textit{a priori}. Lately, when estimating production technologies from the data, it has increasingly become a norm to impose theoretical regularity conditions onto the estimand to ensure that the estimates are structurally meaningful. In the case of a SAR production function, the restriction in \eqref{eq:Qconstr} would also need to be imposed. While global in nature, this constraint depends on data, is highly nonlinear and involves the inversion of an $n\times n$ matrix which can be quite computationally demanding even for moderate sample sizes \citep[see][]{lesagepace}:
\begin{equation}
    \mathbf{i}_n \le \text{diag}\left\{\left[\mathbf{I}_n-\rho\mathbf{D}\right]^{-1}\right\} \le \frac{1}{\beta_K}\mathbf{i}_n.
\end{equation}

\medskip

\noindent\textbf{Identification Considerations.}\textemdash A more fundamental issue with the SAR production-function models concerns their (un)identifiability from the firm-level data due to the endogeneity of the firm's allocations of own variable inputs. The latter can be equivalently characterized as the omitted variable problem, whereby variable inputs are correlated with firm unobservables which we capture using the persistent productivity term $\omega_{it}$. Many studies considering SAR production-function models (or their Durbin extensions) leave this well-known problem unaddressed, focusing only on the endogeneity of the SAR lag term while implausibly assuming that all input regressors are exogenous or that unobservables\textemdash firm productivity, or efficiency\textemdash are purely random \citep[see][]{glassetal2016,glassetal2019,glassetal2020b}. 
Others tackle this problem under the assumption that the endogeneity-inducing correlated unobservables are time-invariant and can be controlled for via firm fixed effects \citep[e.g.,][]{glassetal2016oep,glassetal2020}. 

While clearly an improvement over the random treatment of firm unobservables, the fixed-effect approach however tends to be quite unsatisfactory in practice because differencing or within-transforming of the data necessary to purge fixed effects oftentimes leaves little usable identifying variation which yields unrealistically small and statistically insignificant estimates of the capital elasticity \citep[see][]{gm1998,abbp2007,gnr2013}. Consequently, practitioners have favored tackling endogeneity in the production-function estimation via identification schemes involving proxy variables or external instruments. 

The instrument-based identification of SAR production functions has been considered by \citet{kutluetal2020}, although they provide no guidance about the candidates for valid external instruments. In the stochastic-frontier productivity literature, the instrumentation is most times done using  firm-level variation in prices.\footnote{The less common instruments include demand shifters or the external ``determinants'' of firm productivity/efficiency.} However, the validity and practicality of using lagged firm-level prices for identification is not universal. Not only are the price data often unavailable or prone to measurement errors \citep{lp2003}, but the use of prices may also be problematic on theoretical grounds \citep[see][]{gm1998,abbp2007,acf2015,flynnetal2019}. 
Specifically, the validity of prices as exogenous instruments is normally justified by invoking the assumption of perfectly competitive markets. However, if firms were indeed price-takers, in theory, one should not observe the firm-level variation in prices and, without such a variation, prices cannot be used as operational instruments. Even with the aggregate prices varying exogenously across space, such a variation may be insufficient for identification as shown by \citet{gnr2013}. If a researcher does observe the variation in prices across all individual firms, the latter variation may be reflecting differences in firms' market power and/or the quality of inputs/outputs. For instance, if the firm-level variation in input prices reflects differential quality in inputs, then random updates in prices that render prices valid instruments are likely related to productivity innovations because a more productive firm is to use more productive, higher-quality inputs \citep{flynnetal2019}. Thus, be it due to the market power or quality differentials, the variation in prices will then be endogenous to firms' decisions and hence cannot help the identification \citep[also see][]{gnr2013}. Furthermore, \textit{lagging} the instruments does not help either. Putting the issue of exogeneity aside, \citet{flynnetal2019} raise concerns about the strong conditions on the evolution processes that must be satisfied for the lagged prices to have any strength as instruments.\footnote{Similar arguments can be made about the demand and productivity shifters.} 

The above underscores the practical appeal of proxy-variable identification strategies that do not require external instruments to identify production technologies. Despite some recent attempts at the proxy-variable identification of SAR production functions \citep[see][]{houetal2020}, below we show that such proxy-variable methodologies originally designed for the estimation of non-spatial production functions generally \textit{cannot} be extended to accommodate their SAR specifications. 

For concreteness, we augment the logged SAR production function in \eqref{eq:sar1} to include more than one input and a random shock and where we also resume time-indexing the variables:
\begin{equation}\label{eq:sar2}
	y_{it} = \rho \sum_{j(\ne i)=1}^n d_{ij} y_{jt} + \beta_K k_{it} + \beta_M m_{it} + \omega_{it} + \eta_{it}.
\end{equation}
For simplicity, here we assume that spatial relationships captured by the weights $\mathbf{D}=\{d_{ij}\}$ are time-invariant.
Akin to \eqref{eq:rflog}, the corresponding reduced form of the $i$th firm production function at time period $t$ is
\begin{equation}\label{eq:sar2_rflog}
	y_{it} = \beta_K\sum_{j=1}^n Q_{ij} k_{jt} + \beta_M\sum_{j=1}^n Q_{ij} m_{jt} +\sum_{j=1}^n Q_{ij}\left[\omega_{jt} +\eta_{jt}\right],
\end{equation}
where $\mathbf{Q}=\{Q_{ij}\}$ is an $n\times n$ time-invariant diagonal block of the $nT\times nT$ spatial multiplier matrix
\begin{equation*}
    \left[\mathbf{I}_{nT}-\rho\mathbf{I}_T\otimes\mathbf{D}\right]^{-1}= \mathbf{I}_T\otimes \overbrace{\left[ \mathbf{I}_n-\rho \mathbf{D}\right]^{-1}}^{\mathbf{Q}}.
\end{equation*}

In line with the standard structural assumptions in the proxy-variable productivity literature (and to echo those we make in Section \ref{sec:model}), we assume that 
(\textsl{i}) $K_{it}$ is dynamically optimized with a delay and subject to the adjustment costs, whereas $M_{it}$ is freely varying and chosen statically, 
(\textsl{ii}) persistent firm productivity follows a controlled (location-homogeneous) first-order Markov process with  transition probability $\mathcal{P}^{\omega}(\omega_{it}|\omega_{it-1},G_{it-1})$, and the random shock $\eta_{it}$ is i.i.d., 
(\textsl{iii}) firms are risk-neutral and seek to maximize a discounted stream of expected life-time profits in perfectly competitive output and factor markets. 
Note that, in the case of a SAR formulation of spatial effects, the extent through which firm location plays a role in the production is via the SAR term $\sum_{j(\ne i)=1}^n d_{ij} y_{jt}$ that shifts the frontier; all production-function parameters as well as the productivity evolution process are location-invariant.

As noted earlier, in order to identify the SAR production function in \eqref{eq:sar2}, one needs to tackle the endogeneity of not only the spatial lag $\sum_{j(\ne i)=1}^n d_{ij} y_{jt}$ (due to the simultaneous ``reflection'') but also both firm inputs that are correlated with \textit{unobservable} firm productivity $\omega_{it}$. While the endogenous $\sum_{j(\ne i)=1}^n d_{ij} y_{jt}$ can be handled fairly easily using internal instruments such as the first- and higher-order spatial lags of neighbors' inputs (per the reduced form in \eqref{eq:sar2_rflog}) as typically done in spatial models, the endogeneity of inputs requires far more finesse. Owing to the already-discussed general lack of external instruments, a popular approach to tackling this omitted variable problem is structural and relies on proxying for the ``omitted'' $\omega_{it}$ using the inverted material demand function. Namely, analogous to the steps we take in Section \ref{sec:identification}, making use of the Markovian nature of firm productivity, we can rewrite the SAR production function \eqref{eq:sar2} as
\begin{equation}\label{eq:sar2_markovsub}
	y_{it} = \rho \sum_{j(\ne i)=1}^n d_{ij} y_{jt} + \beta_K k_{it} + \beta_M m_{it} + h(\omega_{it-1},G_{it-1}) + \zeta_{it}+\eta_{it},
\end{equation}
where $h(\cdot)$ is the conditional mean of $\omega_{it}$ which, if desired, can be assumed to be linear, and $\zeta_{it}$ is a productivity innovation. 

Provided that we can construct an (observable) proxy for $\omega_{it-1}$, $k_{it}$ and $G_{it-1}$ are predetermined and weakly exogenous w.r.t.~$\zeta_{it} + \eta_{it}$ but the freely varying $m_{it}$ is not because it is a function of $\zeta_{it}$ (through $\omega_{it}$ based on which the firm statically chooses materials in  period $t$). Thus, both $\sum_{j(\ne i)=1}^n d_{ij} y_{jt}$ and $m_{it}$ are endogenous. Since the spatial lag is easily instrumentable, let us focus on handling the endogeneity of $m_{it}$. 

As we explain in Section \eqref{sec:identification}, in order to identify the production function in flexible inputs such as $m_{it}$, a solution is to exploit a structural link between the production function and the firm's (static) first-order condition for $m_{it}$, which is also what we do in the first step of our methodology. Thus, the firm's restricted expected profit-maximization problem w.r.t.~the flexible material input subject to the already optimized dynamic input $K_{it}$, productivity $\omega_{it}$ and prices $(P^Y_t,P^M_t)'$ is
\begin{align}\label{eq:profitmax_sar}
\max_{M_{it}}\ P_{t}^Y \exp\left\{\rho \sum_{j(\ne i)=1}^n d_{ij} \mathbb{E}[y_{jt}|\mathcal{I}_{it}]\right\} K_{it}^{\beta_K} M_{it}^{\beta_M} \exp\left\{\omega_{it}\right\}\theta  - P_{t}^M M_{it} .
\end{align}

The optimization problem now also includes the \textit{expected} average neighbor log-output $\sum_{j(\ne i)=1}^n d_{ij} \mathbb{E}[y_{jt}|\mathcal{I}_{it}]$ net of random \textit{ex post} shocks $\{\eta_{jt}\}$ unobservable to firms at the time of making decisions: $y_{jt}=\mathbb{E}[y_{jt}|\mathcal{I}_{it}]+\eta_{jt}\ \forall i,j$. Precisely because of the latter and so long as there are spatial spillovers across firms in that $\rho\ne0$, the firm's first-order condition now accounts for ``feedback effects'' whereby the change in $M_{it}$ affects not only $Y_{it}$ but also\textemdash through spillovers\textemdash neighbors' outputs $\{Y_{jt}\}$ which, in turn, affect firm $i$'s output $Y_{it}$ again. Therefore, the corresponding first-order condition is
\begin{align}\label{eq:foc_sar}
\left(\beta_M+\rho\sum_{j(\ne i)=1}^n d_{ij} \frac{\partial \mathbb{E}[y_{jt}|\mathcal{I}_{it}]}{\partial m_{it}}\right)P_{t}^Y \exp\left\{\rho \sum_{j(\ne i)=1}^n d_{ij} \mathbb{E}[y_{jt}|\mathcal{I}_{it}]\right\} K_{it}^{\beta_K} M_{it}^{\beta_M-1} \exp\left\{\omega_{it}\right\}\theta  = P_{t}^M  .
\end{align}

To arrive at the material share equation, we take the log of \eqref{eq:foc_sar} and subtract the production function in \eqref{eq:sar2}:
\begin{align}
v_{it} &= \ln \left[\left(\beta_M+\rho\sum_{j(\ne i)=1}^n d_{ij} {\frac{\partial \mathbb{E}[y_{jt}|\mathcal{I}_{it}]}{\partial m_{it}}}\right)\theta\right] -\rho\sum_{j(\ne i)=1}^n d_{ij}\eta_{jt} - \eta_{it} \notag \\
&= \ln \left[\beta_M\left(1+\rho\sum_{j(\ne i)=1}^n d_{ij} {Q_{ji}}\right)\theta\right] \underbrace{-\rho\sum_{j(\ne i)=1}^n d_{ij}\eta_{jt} - \eta_{it}}_{\epsilon_{it}},\label{eq:fst_sar} 
\end{align}
where we have made a substitution in the second line using the partial $\frac{\partial \mathbb{E}[y_{jt}|\mathcal{I}_{it}]}{\partial m_{it}}=Q_{ji}\beta_M$ obtained from the reduced form in \eqref{eq:sar2_rflog} and, just like before, $v_{it} = \ln \left(P_{t}^M M_{it}\right)-\ln \left( P_{t}^Y Y_{it}\right)$ is the (observable) log nominal share of material costs in total revenue.
Note that, unlike in a model without spatial lag, the composite error term $\epsilon_{it}\equiv-\big(\rho\sum_{j(\ne i)=1}^n d_{ij}\eta_{jt} + \eta_{it}\big)$ in \eqref{eq:fst_sar} follows a spatial moving average process. Although $\epsilon_{it}$ is spatially correlated, this has no impact on identification because, by assumption, shocks $\{\eta_{it}\}$ are all i.i.d. Further note that each $Q_{ji}$ element in \eqref{eq:fst_sar} is a function of spatial weights and the still-unknown $\rho$. To make this more explicit, we write \eqref{eq:fst_sar} as 
\begin{align}\label{eq:fst_sar*}
v_{it} &= \ln \left[\beta_M\left(1+\rho\mathbf{D}_{(i)}\left(\left[ \mathbf{I}_n-\rho \mathbf{D}\right]^{-1}\right)_{(j)}\right)\theta\right] +{\epsilon_{it}}  ,
\end{align}
where $\mathbf{A}_{(i)}$ and $\mathbf{A}_{(j)}$ respectively denote the $i$th row and $j$th column of some matrix $\mathbf{A}$ (and recall that $d_{ij}=0\ \forall i=j$).

The material share equation \eqref{eq:fst_sar} is a nonlinear regression containing \textit{no} endogenous covariates. It therefore might appear at first that it can be seamlessly estimated via nonlinear least squares. However, the parameters in this regression $\boldsymbol{\alpha}\equiv(\beta_M,\rho,\theta)'$ are \textit{not} identified. To make this unidentification more apparent, we recast the nonlinear least-squares estimator of \eqref{eq:fst_sar*} in a GMM framework. 

Namely, we write the nonlinear equation \eqref{eq:fst_sar*} as $v_{it} = h_{it}(\mathbf{D},\boldsymbol{\alpha}) + \epsilon_{it}$, where $h_{it}(\cdot)$ is the regression function. Consider now the identification of $\boldsymbol{\alpha}$ in the following just-identified nonlinear GMM problem that is equivalent to the nonlinear least-squares estimation: 
\begin{align}\label{eq:fst_ident_gmm}
\boldsymbol{\alpha}_0 = \arg\min_{\boldsymbol{\alpha}}\ 
\mathbb{E}\left[\frac{\partial h_{it}(\mathbf{D},\boldsymbol{\alpha})}{\partial\boldsymbol{\alpha}}\Big(v_{it}-h_{it}(\mathbf{D},\boldsymbol{\alpha})\Big)\right]'
\mathbf{W}
\mathbb{E}\left[\frac{\partial h_{it}(\mathbf{D},\boldsymbol{\alpha})}{\partial\boldsymbol{\alpha}}\Big(v_{it}-h_{it}(\mathbf{D},\boldsymbol{\alpha})\Big)\right],
\end{align}
where 
\begin{align}\label{eq:grad}
\frac{\partial h_{it}(\mathbf{D},\boldsymbol{\alpha})}{\partial\boldsymbol{\alpha}} =
\begin{bmatrix}
    {\beta_M}^{-1} \\
    \frac{1}{1+\rho\mathbf{D}_{(i)}\left(\left[ \mathbf{I}_n-\rho \mathbf{D}\right]^{-1}\right)_{(j)}}
    \left( \mathbf{D}_{(i)}\left(\left[ \mathbf{I}_n-\rho \mathbf{D}\right]^{-1}\right)_{(j)} + \rho\mathbf{D}_{(i)}\left(\left[ \mathbf{I}_n-\rho \mathbf{D}\right]^{-1}\mathbf{D}\left[ \mathbf{I}_n-\rho \mathbf{D}\right]^{-1}\right)_{(j)} \right) \\
    {\theta}^{-1}
\end{bmatrix},
\end{align}
and $\mathbf{W}$ is a symmetric positive-definite moment-weighting matrix. 

To see that \eqref{eq:fst_ident_gmm}--\eqref{eq:grad} do \textit{not} identify all of the $\boldsymbol{\alpha}$ parameters, letting the element in the second row of $\frac{\partial h_{it}(\mathbf{D},\boldsymbol{\alpha})}{\partial\boldsymbol{\alpha}}$ be pictorially denoted by ``$\Box$,'' consider the corresponding information matrix:
\begin{align}\label{eq:fst_ident_gmm_info}
\Psi(\boldsymbol\alpha) 
&= \mathbb{E}\left[\frac{\partial h_{it}(\mathbf{D},\boldsymbol{\alpha})}{\partial\boldsymbol{\alpha}} \frac{\partial h_{it}(\mathbf{D},\boldsymbol{\alpha})}{\partial\boldsymbol{\alpha}'} \right] =
\mathbb{E}\left[
\begin{matrix}
    \frac{1}{\beta_M^2} & \frac{1}{\beta_M}\Box & \frac{1}{\beta_M\theta} \\
    \frac{1}{\beta_M}\Box & \Box^2 & \frac{1}{\theta}\Box \\
    \frac{1}{\beta_M\theta} & \frac{1}{\theta}\Box & \frac{1}{\theta^2}
\end{matrix}\right].
\end{align}

The above $3\times 3$ matrix $\Psi(\boldsymbol\alpha) $ has rank of 1. Thus, the information matrix for the GMM problem in \eqref{eq:fst_ident_gmm} when evaluated at the true parameter values $\Psi(\boldsymbol{\alpha}_0)$ will be rank-deficient, and the parameters in $\boldsymbol{\alpha}$ will therefore be {un}identified \citep[see][]{rothenberg1971}. 

It is also important to note that augmenting the nonlinear least-squares moment restrictions with the unconditional moment corresponding to $\theta$ given in \eqref{eq:theta} analogously to what we do in the first step of our identification methodology will \textit{not} remedy the unidentification problem. More concretely, recalling that $\theta\equiv \mathbb{E}\left[\exp\left\{ \eta_{it}\right\} \right]$ and inverting the composite error $\epsilon_{it}$ appearing in \eqref{eq:fst_sar*} from the spatial moving average process that it follows, we have the additional GMM moment restriction:
\begin{align}\label{eq:fst_ident_gmm_add}
0 = \mathbb{E}\left[g_{it}(\mathbf{D},\boldsymbol{\alpha}) \right] &\equiv \mathbb{E}\left[\exp\left\{ \eta_{it}\right\}-\theta \right] \notag \\
&= \mathbb{E}\left[\exp\left\{ -\left(\left[\mathbf{I}_n+\rho\mathbf{D}\right]^{-1}\right)_{(i)}\boldsymbol{\epsilon}_t\right\}-\theta \right],
\end{align}
where $\boldsymbol{\epsilon}_t=(\epsilon_{1t},\dots,\epsilon_{nt})'$ and each $\epsilon_{it}=v_{it}-h_{it}(\mathbf{D},\boldsymbol{\alpha})=v_{it}-\ln \left[\beta_M\left(1+\rho\mathbf{D}_{(i)}\left(\left[ \mathbf{I}_n-\rho \mathbf{D}\right]^{-1}\right)_{(j)}\right)\theta\right]$. With this, let us consider the information matrix  $I(\boldsymbol{\alpha})=[\Psi(\boldsymbol{\alpha}), \Phi(\boldsymbol{\alpha})]$ for an augmented GMM problem, where 
\begin{align}
\Phi(\boldsymbol{\alpha})=\mathbb{E}\left[\frac{\partial g_{it}(\mathbf{D},\boldsymbol{\alpha})}{\partial\boldsymbol{\alpha}}\right] =
\mathbb{E}\left[ 
\begin{matrix}
    \beta_M^{-1}\exp\left\{ -\left(\left[\mathbf{I}_n+\rho\mathbf{D}\right]^{-1}\right)_{(i)}\boldsymbol{\epsilon}_t\right\}\left(\left[\mathbf{I}_n+\rho\mathbf{D}\right]^{-1}\right)_{(i)}\mathbf{i}_n \\
    \frac{\partial g_{it}(\mathbf{D},\boldsymbol{\alpha})}{\partial\rho} \\
    \theta^{-1}\exp\left\{ -\left(\left[\mathbf{I}_n+\rho\mathbf{D}\right]^{-1}\right)_{(i)}\boldsymbol{\epsilon}_t\right\}\left(\left[\mathbf{I}_n+\rho\mathbf{D}\right]^{-1}\right)_{(i)}\mathbf{i}_n-1
\end{matrix}\right].
\end{align}

It is apparent that the information matrix $I(\boldsymbol{\alpha})$ even when including the additional moment is still not full-rank, and the material share equation remains unidentified. 

We have thus shown that, notwithstanding the \citet{gnr2013} result for non-spatial proxy-variable estimators, the SAR production function cannot be identified in flexible inputs by exploiting a structural link between the production function and the firm's (static) optimality conditions. Consequently, the whole model is unidentified. 
This is in stark contrast with the main result of our paper. Our ability to capture locational effects in production and achieve proxy-variable identification of the production technology and firm productivity in the presence of technology spillovers and agglomeration economies stems from our fundamentally different conceptualization of cross-firm spatial interdependence. Our methodology incorporates firm location through local smoothing, which models the production technology for each location as the geographically weighted average of the input-output \textit{relationships} for firms in the nearby locations, whereas the SAR production-function models formulate locational aspects using the spatially weighed averages of the output/inputs \textit{quantities} while keeping the production technology location-invariant.

}


\section{Translog Technology}
\label{sec:appx_tl}

Our methodology can adopt more flexible specifications of the firm's production technology. The log-quadratic translog specification provides a natural extension of the log-linear Cobb-Douglas form. See \citet{deloeckerwarzynski2012} and \citet{deloeckeretal2016} for recent applications of the translog production functions in the structural proxy estimation. Just like we have done in  \eqref{eq:prodfn_hicks_cd}, we generalize the standard fixed-parameter translog specification to accommodate potential locational heterogeneity in production by letting its coefficients vary with the firm's location in a nonparametric way, i.e.,
\begin{align}\label{eq:prodfn_hicks_tl}
\ln F_{|S_i}(\cdot) =&\ \beta_K(S_i)k_{it}+\tfrac{1}{2}\beta_{KK}(S_i)k_{it}^2+ \beta_L(S_i)l_{it}+\tfrac{1}{2}\beta_{LL}(S_i)l_{it}^2 +\beta_M(S_i)m_{it}+\tfrac{1}{2}\beta_{MM}(S_i)m_{it}^2\ + \notag \\ 
&\ \beta_{KL}(S_i)k_{it}l_{it}+\beta_{KM}(S_i)k_{it}m_{it}+\beta_{LM}(S_i)l_{it}m_{it} .
\end{align}

Our methodology can then be modified as follows.

\textsl{First step}.\textemdash The firm's first-order condition for the static optimization in \eqref{eq:profitmax} with respect to $M_{it}$ is now takes the following form:
\begin{equation}\label{eq:foc_tl}
\ln P_{t}^Y+\ln F_{|S_i} + \ln\left[ \beta_M(S_i)+\beta_{MM}(S_i)m_{it}+\beta_{KM}(S_i)k_{it}+\beta_{LM}(S_i)l_{it} \right] - m_{it} + \omega_{it}+ \ln \theta = \ln P_{t}^M ,
\end{equation}
where $\ln F_{|S_i}$ equals the semiparametric translog technology in \eqref{eq:prodfn_hicks_tl}. The location-specific material share equation corresponding to this optimality condition is now given by
\begin{equation}\label{eq:fst_tl}
v_{it} = \ln \Big(\Big[\beta_M(S_i)+\beta_{MM}(S_i)m_{it}+\beta_{KM}(S_i)k_{it}+\beta_{LM}(S_i)l_{it} \Big]\theta\Big) - \eta_{it} ,
\end{equation}
where $\beta_M(S_i)+\beta_{MM}(S_i)m_{it}+\beta_{KM}(S_i)k_{it}+\beta_{LM}(S_i)l_{it}$ is the material elasticity function. Analogous to the discussion in Section \ref{sec:identification}, the above log material share equation identifies the locationally varying material-related production-function parameters $(\beta_M(S_i),\beta_{MM}(S_i),\beta_{KM}(S_i),\beta_{LM}(S_i))'$ as well as the mean of exponentiated shocks $\theta$ based on the mean-orthogonality condition $\mathbb{E}[\eta_{it} | \mathcal{I}_{it}] =\mathbb{E}[\eta_{it} | k_{it}, l_{it}, m_{it}, S_i]= \mathbb{E}[\eta_{it}] = 0$. 

\textsl{Second step}.\textemdash Having identified the production technology in the dimension of its endogenous freely varying input $M_{it}$, we can focus on the rest of production function. With the already identified $y_{it}^*\equiv y_{it} - \beta_M(S_i)m_{it}-\tfrac{1}{2}\beta_{MM}(S_i)m_{it}^2-\beta_{KM}(S_i)k_{it}m_{it}-\beta_{LM}(S_i)l_{it}m_{it}$ and using the inverted conditional material demand derived from \eqref{eq:foc_tl} to substitute for $\omega_{it-1}$, we now have the analogue of \eqref{eq:sst}:
\begin{align}\label{eq:sst_tl}
y_{it}^* =&\ \beta_K(S_i)k_{it}+\tfrac{1}{2}\beta_{KK}(S_i)k_{it}^2+ \beta_L(S_i)l_{it}+\tfrac{1}{2}\beta_{LL}(S_i)l_{it}^2 + \beta_{KL}(S_i)k_{it}l_{it}+\rho_0(S_i)\ + \notag  \\
&\ \rho_1(S_i) \Big[\nu^*_{it-1}- \beta_K(S_i)k_{i,t-1}-\tfrac{1}{2}\beta_{KK}(S_i)k_{i,t-1}^2- \beta_L(S_i)l_{i,t-1}-\tfrac{1}{2}\beta_{LL}(S_i)l_{i,t-1}^2 - \beta_{KL}(S_i)k_{i,t-1}l_{i,t-1}\Big]\ + \notag \\
&\ \rho_2(S_i)G_{it-1} +  \zeta_{it} + \eta_{it},
\end{align}
where 
\begin{align*}
\nu_{it-1}^*=&\ \ln[ P_{t-1}^M/P_{t-1}^Y]-\ln (\left[\beta_M(S_i)+\beta_{MM}(S_i)m_{it-1}+\beta_{KM}(S_i)k_{it-1}+\beta_{LM}(S_i)l_{it-1} \right]\theta)\ + \\ 
&\ [1-\beta_M(S_i)]m_{it-1}-\tfrac{1}{2}\beta_{MM}(S_i)m_{it-1}^2-\beta_{KM}(S_i)k_{it-1}m_{it-1}-\beta_{LM}(S_i)l_{it-1}m_{it-1}
\end{align*}
is already identified/observable and predetermined with respect to $\zeta_{it}+ \eta_{it}$. All covariates in \eqref{eq:sst_tl} are predetermined and can self-instrument thereby identifying the translog model.

\medskip

\noindent \textbf{Estimation.}\textemdash The estimation methodology here mirrors that used for the Cobb-Douglas model except that the local-constant least-squares estimator in the first step is now also nonlinear. 

Assuming that all unknown locationally varying coefficient functions are smooth and twice continuously differentiable in the neighborhood of $S_i=s$, the location-specific material share equation in \eqref{eq:fst_tl} can be locally approximated around $s$ via constants as
\begin{align}\label{eq:fst_tl_est}
v_{it} &\approx \ln \Big(\Big[\beta_M(s)+\beta_{MM}(s)m_{it}+\beta_{KM}(s)k_{it}+\beta_{LM}(s)l_{it} \Big]\theta\Big) - \eta_{it} \notag \\
&\approx \ln \Big([\beta_M(s)\theta]+[\beta_{MM}(s)\theta]m_{it}+[\beta_{KM}(s)\theta]k_{it}+[\beta_{LM}(s)\theta]l_{it}\Big) - \eta_{it},
\end{align}
with the corresponding kernel estimator of $\Theta_1(s)=[\beta_M(s)\theta,\beta_{MM}(s)\theta,\beta_{KM}(s)\theta,\beta_{LM}(s)\theta]'$ being
\begin{align}
\widehat{\Theta}_1(s) = \arg\min_{\Theta_1(s)}\ \sum_i\sum_t &\ \mathcal{K}_{h_1}(S_i,s)\Big( v_{it} - \ln \Big([\beta_M(s)\theta]+[\beta_{MM}(s)\theta]m_{it}+[\beta_{KM}(s)\theta]k_{it}+[\beta_{LM}(s)\theta]l_{it}\Big) \Big)^2.
\end{align}

To estimate the material elasticity parameter functions  $[\beta_M(s),\beta_{MM}(s),\beta_{KM}(s),\beta_{LM}(s)]'$ around $S_i=s$ net of constant $\theta$, we first recover $\widehat{\theta}$ as
\begin{equation}
\widehat{\theta} = \sum_i\sum_t\exp\left\{ \ln \Big([\widehat{\beta_M(s)\theta}]+[\widehat{\beta_{MM}(s)\theta}]m_{it}+[\widehat{\beta_{KM}(s)\theta}]k_{it}+[\widehat{\beta_{LM}(s)\theta}]l_{it}\Big)-v_{it}\right\}
\end{equation}
and then use it to scale $\widehat{\Theta}_1(s)$ which yields our first-step estimator:
\begin{equation}
\left[\widehat{\beta}_M(s),\widehat{\beta}_{MM}(s),\widehat{\beta}_{KM}(s),\widehat{\beta}_{LM}(s)\right]' = nT \widehat{\Theta}_1(s)\Big/\widehat{\theta}.
\end{equation}

Using these first-step local estimates, we then construct $\widehat{y}_{it}^*= y_{it} - \widehat{\beta}_M(S_i)m_{it}-\tfrac{1}{2}\widehat{\beta}_{MM}(S_i)m_{it}^2-\widehat{\beta}_{KM}(S_i)k_{it}m_{it}-\widehat{\beta}_{LM}(S_i)l_{it}m_{it}$ and $\widehat{\nu}^*_{it-1}=\ln[ P_{t-1}^M/P_{t-1}^Y]-\ln ([\widehat{\beta}_M(S_i)+\beta_{MM}(S_i)m_{it-1}+\widehat{\beta}_{KM}(S_i)k_{it-1}+\widehat{\beta}_{LM}(S_i)l_{it-1} ]\theta)+ [1-\widehat{\beta}_M(S_i)]m_{it-1}-\tfrac{1}{2}\widehat{\beta}_{MM}(S_i)m_{it-1}^2-\widehat{\beta}_{KM}(S_i)k_{it-1}m_{it-1}-\widehat{\beta}_{LM}(S_i)l_{it-1}m_{it-1}$. 

Analogous to the first-step estimation, we then locally approximate each unknown parameter function in \eqref{eq:sst_tl} via local-constant approach. Collectively denoting all unknown parameters in the equation as $\Theta_2(S_i)=[\beta_K(S_i),\beta_{KK}(S_i),\beta_L(S_i),\beta_{LL}(S_i),\beta_{KL}(S_i),\rho_0(S_i),\rho_1(S_i),\rho_2(S_i)]'$, the second-step local-constant nonlinear least-squares estimator in the neighborhood of $S_i=s$ is then given by
\begin{align}\label{eq:sst_tl_est}
\widehat{\Theta}_2(s) = &\ \arg\min_{\Theta_2(s)}\  \sum_i\sum_t  \mathcal{K}_{h_2}(S_i,s)\Big( y_{it}^* - \beta_K(S_i)k_{it}-\tfrac{1}{2}\beta_{KK}(S_i)k_{it}^2- \beta_L(S_i)l_{it}-\tfrac{1}{2}\beta_{LL}(S_i)l_{it}^2 - \beta_{KL}(S_i)k_{it}l_{it}\ - \notag \\ &\ \rho_1(S_i) \Big[\nu^*_{it-1}- \beta_K(S_i)k_{i,t-1}-\tfrac{1}{2}\beta_{KK}(S_i)k_{i,t-1}^2- \beta_L(S_i)l_{i,t-1}-\tfrac{1}{2}\beta_{LL}(S_i)l_{i,t-1}^2 - \beta_{KL}(S_i)k_{i,t-1}l_{i,t-1}\Big]\ - \notag \\
&\  \rho_0(S_i) - \rho_2(S_i)G_{it-1} \Big)^2.
\end{align}

{
\section{Finite-Sample Performance of the Estimator}
\label{sec:appex_sim}
\setcounter{table}{0}
\setcounter{figure}{0}

We investigate the finite-sample performance of our proposed estimation procedure in a set of Monte Carlo experiments. Our data generating process (DGP) builds on the setup in \citet{griecoetal2016} and \citet{malikovetal2020} which we modify to allow for locational heterogeneity.

Without loss of generality, we dispense with labor and consider the production process with two inputs only: a quasi-fixed capital and freely varying materials. Let the true technology take a semiparametric locationally-varying Cobb-Douglas form:
\begin{equation}
	Y_{it} = K_{it}^{\beta_K(S_i)}M_{it}^{\beta_M(S_i)}\exp\{\omega_{it}\}\exp\{\eta_{it}\}.
\end{equation}

To simplify matters, we assume that all firms are located on a straight line, with the univariate location variable $S_i$ indexing their relative location. We assume a discrete uniform spatial distribution of firms, with $S_i\in \mathbb{S}=\{0.50,0.51,\dots,0.98,0.99 \}$ and $D=50$ locations. The random disturbance is $\eta_{it}\sim\text{i.i.d.}\ \mathbb{N}(0,0.07^2)$.

We assume the decreasing returns to scale across all locations but let the scale elasticity differ across locations. Concretely, we have that the returns to scale increase as one moves rightwards in space $\mathbb{S}$ by having the two input elasticity functions smoothly vary across locations as follows:
\begin{align}
	\beta_K(S_i)&=0.2 + 0.1 S_i\\
	\beta_M(S_i)&=0.4 + 0.1\exp(S_i^2).
\end{align}

The productivity components are generated as follows. We model the persistent productivity as a location-specific exogenous AR(1) process:
\begin{align}
	\omega_{it}=\rho_0(S_i)+\rho_1(S_i)\omega_{it-1}+\zeta_{it},
\end{align}
where we set $\rho_0(S_i)=0.5+S_i-S_i^2$ and $\rho_1(S_i)=0.7 \ \forall\ S_i$. In this, we assume that the mean firm productivity is the highest in the middle of $\mathbb{S}$ and symmetrically diminishes in both direction therefrom. The firm's initial level of productivity $\omega_{i1}$ is set to $\rho_0(S_i)$ and therefore is determined purely by the firm's location. The productivity innovation is $\zeta_{it}\sim\text{i.i.d.}\ \mathbb{N}(0,0.04^2)$.

The firm's capital is set to evolve according to $K_{it}= I_{it-1}+(1-\delta_i)K_{it-1}$, with the firm-specific depreciation rates $\delta_i$ uniformly drawn from $\{0.05,0.075,0.10,0.125,0.15\}$. The initial levels of capital $K_{i0}$ is drawn from $\mathbb{U}(10,200)$ identically and independently distributed over $i$. The investment function takes the following form: $I_{it-1}= K_{it-1}^{\alpha_1} \exp\{\alpha_2\omega_{it-1}\}$, where $\alpha_1=0.8$ and $\alpha_2=0.1$.

The materials $M_{it}$ series is generated solving the firm's restricted expected profit maximization problem along the lines of \eqref{eq:profitmax}. The conditional demand for $M_{it}$ is given by
\begin{align}
	M_{it}&= \arg\max_{\mathcal{M}_{it}}\ \Big\{ P_{t}^Y K_{it}^{\beta_K(S_i)}\mathcal{M}_{it}^{\beta_M(S_i)} \exp\{\omega_{it}\}\theta  - P_{t}^M \mathcal{M}_{it} \Big\}  
	=\left[ \beta_M(S_i) K_{it}^{\beta_K(S_i)}\exp\{\omega_{it}\}\right]^{1/(1-\beta_M(S_i))},
\end{align}
where, in the second equality, we have normalized $P_t^M=\theta\ \forall\ t$ and have assumed no temporal variation in output prices: $P^{Y}_{t}=1$ for all $t$.

\begin{table}[t]
	\caption{Second-Step Estimates of Locationally-Varying Parameters}\label{tab:sim}
	\centering
	\small
	\begin{tabular}{l rrr | rrr }
		\toprule[1pt]
		& \multicolumn{3}{c}{Panel A: Kernel-Smoothing} 
		& \multicolumn{3}{c}{Panel B: Sample-Splitting}  \\[2pt]
		& $\beta_K(\cdot)$ & $\rho_1(\cdot)$ & $\rho_2(\cdot)$ 
		& $\beta_K(\cdot)$ & $\rho_1(\cdot)$ & $\rho_2(\cdot)$ \\
		\midrule
		
		&\multicolumn{6}{c}{$n=100$}  \\[2pt]
		Mean Bias 		&--0.0113& 0.0065 & 0.0046 &--0.0059&--0.0231& --0.0297 \\
		RMSE 			& 0.0569 & 0.0731 & 0.0329 & 0.1725 & 0.2552 & 0.1152 \\
		MAE 			& 0.0449 & 0.0593 & 0.0267 & 0.1446 & 0.1991 & 0.0885 \\[4pt]
		
		&\multicolumn{6}{c}{$n=200$}  \\[2pt]
		Mean Bias 		&--0.0069& 0.0037 & 0.0036 &--0.0117& 0.0016 & --0.0085 \\
		RMSE 			& 0.0388 & 0.0508 & 0.0238 & 0.1196 & 0.1577 & 0.0693 \\
		MAE 			& 0.0311 & 0.0413 & 0.0193 & 0.0937 & 0.1239 & 0.0546 \\[4pt]
		
		&\multicolumn{6}{c}{$n=400$}  \\[2pt]
		Mean Bias 		&--0.0070& 0.0030 & 0.0053 &--0.0073& 0.0034 & --0.0028 \\
		RMSE 			& 0.0278 & 0.0356 & 0.0178 & 0.0797 & 0.1027 & 0.0456 \\
		MAE 			& 0.0223 & 0.0290 & 0.0144 & 0.0608 & 0.0808 & 0.0362 \\[4pt]
		
		\midrule   
		\multicolumn{7}{p{10.5cm}}{\scriptsize Ours is a kernel-smoothing estimator which uses information from \textit{all} locations, albeit weighting it based on the proximity to a location of interest. The sample-splitting estimator is essentially a ``frequency estimator'' which splits the data sample by location to estimates location-specific parameters using information from that location only.  $T=10$ throughout.} \\
		\bottomrule[1pt] 
	\end{tabular}
\end{table}

We estimate the model via the two-step kernel-smoothing estimation algorithm outlined in Section \ref{sec:estimation}. Although we cross-validate the optimal number of nearest neighbors ($h$) in the empirical application, to conserve computational time, in our simulations we rely on the result that the optimal $h$ when cross-validating is $h\propto n^{4/(4+\dim(S_i))}$  \citep[see][]{ouyangetal2006} and set $h=0.3(nT)^{4/5}$. We consider a balanced panel of $n=\{100,200,400\}$ firms operating during $T=10$ periods. Each panel is simulated $Q=500$ times. For each simulation repetition, we compute the mean bias, the root mean squared error (RMSE) and the mean absolute error (MAE) over all firms. Panel A of Table \ref{tab:sim} reports these metrics averaged across $Q$ simulations for the capital elasticity $\beta_K(S_i)$ and the productivity parameters $\rho_0(S_i)$ and $\rho_1(S_i)$.\footnote{We omit the results corresponding to the material elasticity $\beta_M(S_i)$ estimated in the first step because the estimator yields very precise estimates even for small sample sizes.}

The simulation results for our estimator are encouraging and show that our methodology recovers the true parameters remarkably well, thereby lending strong support to the validity of our identification strategy. As expected of a consistent estimator, the estimation becomes more stable as $n$ grows. Furthermore, our estimator significantly outperforms a crude\textemdash albeit computationally simpler\textemdash alternative estimator which splits the data sample by location to (parametrically) estimates location-specific parameters using the information from that location only. The results for this sample-splitting estimator are summarized in Panel B of Table \ref{tab:sim}. Such an alternative estimation procedure is also less practical because its feasibility is dependent on having enough data from each unique location. Our estimation procedure is immune to this problem because it uses information from \textit{all} locations but with varying degree of relative importance as determined by their proximity to a location of interest.}


\section{Bootstrap Inference}
\label{sec:appx_inference}
\setcounter{table}{0}
\setcounter{figure}{0}

Due to a multi-step nature of our estimator as well as the presence of nonparametric components, computation of the asymptotic variance of the estimators is not simple. {For statistical inference, we therefore use bootstrap.} We approximate sampling distributions of the estimators via wild residual block bootstrap that takes into account a panel structure of the data, with all the steps bootstrapped jointly owing to a sequential nature of our estimation procedure.
Concretely, the bootstrap algorithm is as follows.
\begin{enumerate}\itemsep 0pt
	\item Compute the two steps of our estimation procedure using the original data. Denote the obtained estimates as $\big(\widehat{\beta_M}(S_i),\widehat{\theta},\widehat{\Theta}(S_i)'\big)'$ for all $i=1,\dots,n$. Let the (negative of) first-step residuals be $\{\widehat{\eta}_{it}\}$ and the second-step residuals be $\{\widehat{ \zeta_{it}+ \eta_{it}}\}$. Recenter these.
	
	\item Generate bootstrap weights $\xi_{i}^b$ for all cross-sectional units $i=1,\dots,n$ from the \citet{mammen1993} two-point mass distribution:
	\begin{eqnarray}\label{eq:two_point}
		\xi_{i}^b=%
		\begin{cases}
			\frac{1+\sqrt{5}}{2} & \text{with prob.}\quad \frac{\sqrt{5}-1}{2\sqrt{5}} \\
			\frac{1-\sqrt{5}}{2} & \text{with prob.}\quad \frac{\sqrt{5}+1}{2\sqrt{5}}.
		\end{cases}
	\end{eqnarray}
	Next, for each observation $(i,t)$ with $i=1,\dots,n$ and $t=1,\dots,T$, jointly generate a new bootstrap first-step disturbance $\eta_{it}^b = \xi_{i}^b\times\widehat{\eta}_{it}$ and a new bootstrap second-step disturbance $(\zeta_{it}+ \eta_{it})^b=\xi_{i}^b\times(\widehat{ \zeta_{it}+ \eta_{it}})$.
	
	\item Generate a new bootstrap first-step outcome variable via $v_{it}^b=\ln \left[\widehat{\beta}_M(S_i)\widehat{\theta}\right]-\eta_{it}^b$ for all $i=1,\dots,n$ and $t=1,\dots,T$. 
	
	\item Generate a new bootstrap second-step outcome variable using $y_{it}^{*b}=\widehat{\beta}_K(S_i) k_{it}+\widehat{\beta}_L(S_i)l_{it}+ \widehat{\rho}_0(S_i)+ \widehat{\rho}_1(S_i) \Big[\widehat{\nu}^{*}_{it-1}-\widehat{\beta}_K(S_i) k_{it-1}-\widehat{\beta}_L(S_i)l_{it-1}\Big]+\widehat{\rho}_2(S_i)G_{it-1}+(\zeta_{it}+ \eta_{it})^b$ for all $i=1,\dots,n$ and $t=1,\dots,T$, where $\widehat{\nu}^{*}_{it-1}=\ln[ P_{t-1}^M/P_{t-1}^Y]-\ln [\widehat{\beta}_M(S_i)\widehat{\theta}]+[1-\widehat{\beta}_M(S_i)]m_{it-1}$ is constructed using the original parameter estimates. 

	\item Recompute the first step using $\{v_{it}^b\}$ in place of $\{v_{it}\}$ and, for all $i$, denote the obtained coefficient estimates as $\big(\widehat{\beta}_M^b(S_i),\widehat{\theta}^b\big)'$. Use these bootstrap estimates to construct $\widehat{\nu}^{*b}_{it-1}=\ln[ P_{t-1}^M/P_{t-1}^Y]-\ln [\widehat{\beta}_M^b(S_i)\widehat{\theta}^b]+[1-\widehat{\beta}_M^b(S_i)]m_{it-1}$ for all $i=1,\dots,n$ and $t=1,\dots,T$. 
	
	\item Recompute the second step using $\{y_{it}^{*b}\}$ in place of $\{\widehat{y}_{it}^{*}\}$. When recomputing the models, also use  $\{\widehat{\nu}^{*b}_{it-1}\}$ in place of $\{\widehat{\nu}^{*}_{it-1}\}$. For all $i=1,\dots, n$, denote the obtained coefficient estimates as  $\widehat{\Theta}^b(S_i)=\big(\widehat{\beta}_K^b(S_i),\widehat{\beta}_L^b(S_i),\widehat{\rho}^{\omega b}_0(S_i),\widehat{\rho}^{\omega b}_1(S_i), \widehat{\rho}^{\omega b}_2(S_i)\big)'$.
	
	\item Repeat steps 2 through 6 of the algorithm $B$ times. 
\end{enumerate} 

{Inference is performed using the bootstrap percentile confidence intervals.} Let the observation-specific estimand of interest be denoted by $\mathcal{E}$, e.g., the firm $i$'s labor elasticity coefficient $\beta_L(S_i)$ or the returns to scale defined as the sums of $\beta_K(S_i)$, $\beta_L(S_i)$ and $\beta_M(S_i)$. {To test two-tailed hypotheses, we can use the empirical distribution of $\{\widehat{\mathcal{E}}^1,\dots,\widehat{\mathcal{E}}^B\}$ to estimate the \textit{two}-sided $(1-\alpha)\times100$\% confidence bounds for $\mathcal{E}$ as an interval between the $[\alpha/2\times100]$th and $[(1-\alpha/2)\times100]$th percentiles of the bootstrap distribution. Naturally, for one-tailed hypotheses, to estimate the \textit{one}-sided lower or upper $(1-\alpha)\times100$\% confidence bound, we can use the $[\alpha\times100]$th or $[(1-\alpha)\times100]$th bootstrap percentiles, respectively.

\medskip

\noindent\textbf{Finite-Sample Performance.}\textemdash Using the DGP described in Appendix \ref{sec:appex_sim}, we investigate a finite-sample performance of the above bootstrap procedure in a simulation. This is of interest because of the complexity of our multi-step estimation procedure, which makes establishing the validity of bootstrap nontrivial.
We focus on the two-sided 95\% confidence intervals ($\alpha=0.05$) for (\textsl{i}) the \textit{average} elasticity of capital across all locations $\overline{\beta}_K=\frac{1}{D}\sum_{S_i\in\mathbb{S}}\beta_K(S_i)$ and (\textsl{ii}) capital elasticity at select locations $\beta_K(S_0)$. We choose $S_0=\{0.65,0.75,0.85\}$ which roughly correspond to quartiles of $\mathbb{S}$. To conserve computational time, the number of simulations is $Q = 300$ with $B = 200$ bootstrap replications per each simulation.

\begin{table}[t]
	\caption{Coverage Probability of the Two-Sided 95\% Bootstrap Confidence Interval}\label{tab:bootsim}
	\centering
	\small
	\begin{tabular}{l cccc}
		\toprule[1pt]
		& $\quad \overline{\beta}_K\quad $ & $\beta_K(0.65)$ & $\beta_K(0.75)$ & $\beta_K(0.85)$ \\
		\midrule
		\multicolumn{5}{l}{Panel A: True Parameters are Location-Specific}  \\[2pt]				
		$n=200\qquad$	& 0.873 & 0.923 & 0.901 & 0.850 \\
		$n=400$			& 0.843 & 0.907 & 0.933 & 0.883 \\
		$n=800$			& 0.950 & 0.933 & 0.943 & 0.953 \\[4pt]
		\multicolumn{5}{l}{Panel B: True Parameters are Location-Invariant}  \\[2pt]	
		$n=200$			& 0.937 & 0.943 & 0.943 & 0.950 \\
		$n=400$			& 0.963 & 0.957 & 0.970 & 0.960 \\
		$n=800$			& 0.963 & 0.967 & 0.963 & 0.977 \\
		\midrule   
		\multicolumn{5}{p{8.3cm}}{\scriptsize Panel A (B) corresponds to the DGP with locationally-varying (location-invariant) coefficients. In both cases, the estimation was performed using our kernel-weighting estimator under the presumption that coefficients are location-specific. $T=10$ throughout.} \\
		\bottomrule[1pt] 
	\end{tabular}
\end{table}

\begin{figure}[p]
	\centering
	\includegraphics[scale=0.32]{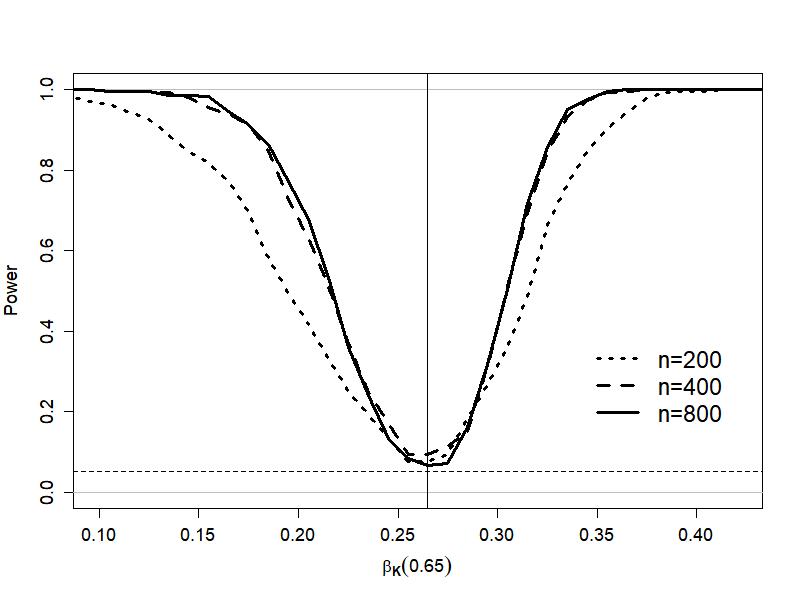}	
	\includegraphics[scale=0.32]{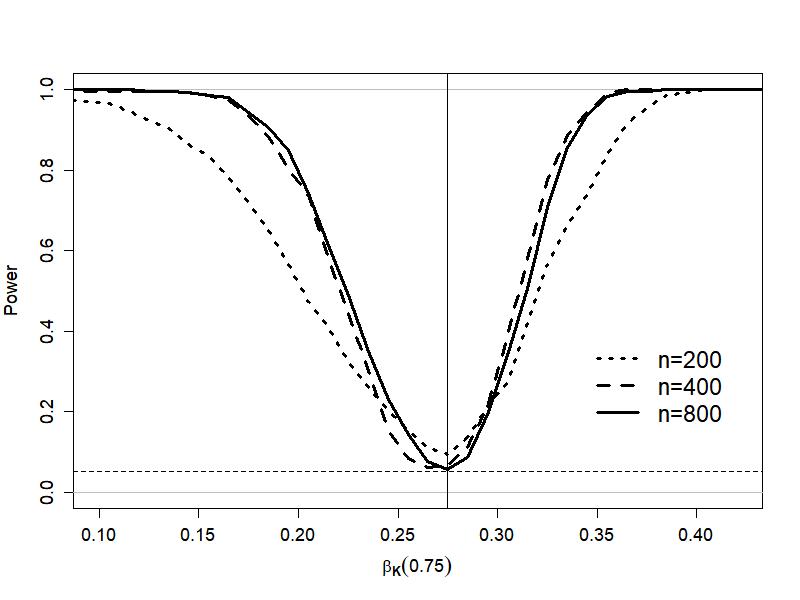}
	\includegraphics[scale=0.32]{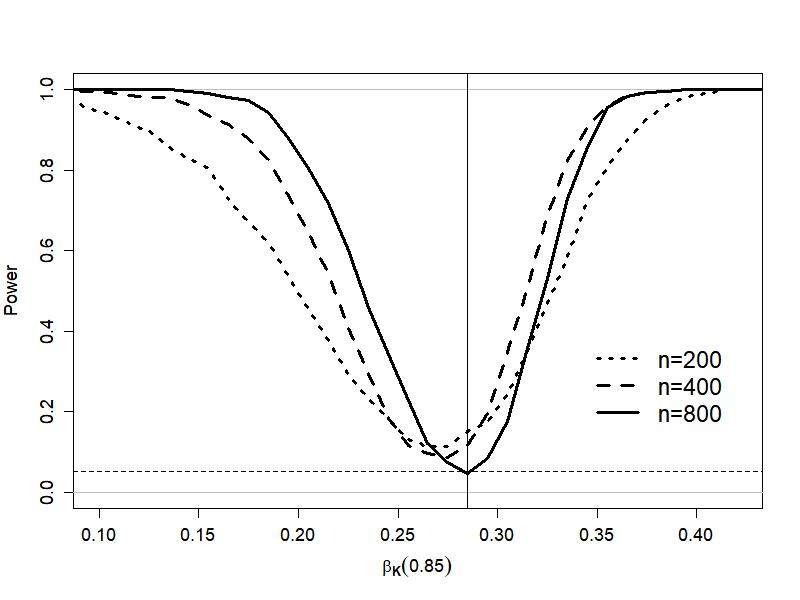}
	\caption{Power of the Two-Sided 95\% Bootstrap Confidence Interval \\ {\scriptsize (Vertical line corresponds to the true $\beta_K(s)$ value)}} \label{fig:boot_power}
\end{figure}

Panel A in Table \ref{tab:bootsim} reports coverage probabilities of the bootstrap confidence intervals for capital elasticity under our DGP for the sample size $n=\{200,400,800\}$. The coverage probability is estimated as the relative frequency (over $Q$ simulations) of the estimated 95\% confidence interval containing the true elasticity value. We also plot power curves for these confidence intervals; see Figure \ref{fig:boot_power}. Here, power is computed as the relative rejection frequency against different nulls on the $x$-axis. The simulations show a satisfactory performance of our bootstrap confidence intervals in finite samples. The results indicate that there may be size distortions for small $n$, which is common for nonparametric tests. However, for a sample size modestly large enough, the estimated coverage is close to the correct coverage. The intervals exhibit good power, which improves as $n$ grows as anticipated of a consistent test.

When the DGP is such that the true parameters are actually location-invariant (i.e., fixed), the performance of bootstrap improves and the coverage probabilities match the nominal confidence level even for small $n$. See Panel B of Table \ref{tab:bootsim}. This is expected because the nonparametric estimation improves significantly when the true process is parametric. }

\medskip

\noindent\textbf{Bias-Corrected Inference.}\textemdash {\citeauthor{efron1982}'s (1982) bias-corrected bootstrap percentile confidence intervals provide means to robustify inference by correcting for the estimator's finite-sample bias. In this case,} the bias-corrected \textit{two}-sided $(1-\alpha)\times100$\% confidence bounds for $\mathcal{E}$ are estimated as an interval between the $[a_1\times100]$th and $[a_2\times100]$th percentiles of the bootstrap distribution, where $a_{1}=\Phi\left(2\widehat{z}_{0}+\Phi^{-1}(\alpha/2)\right) $ and $a_{2}=\Phi\left(2\widehat{z}_{0}+\Phi^{-1}(1-\alpha/2)\right)$ with $\Phi(\cdot)$ being the standard normal cdf along with its quantile function $\Phi^{-1}(\cdot)$. Parameter $ \widehat{z}_0 =  \Phi^{-1}\left(\#\big\{\widehat{\mathcal{E}}^b<\widehat{\mathcal{E}}\big\}/{B}\right)$ is a bias-correction factor measuring median bias, with $\#\{\mathcal{A}\}$ being a count function that returns the number of times event $\mathcal{A}$ is true. Analogously, to estimate the \textit{one}-sided lower or upper $(1-\alpha)\times100$\% confidence bound with bias correction, we can respectively use the $[o_1\times100]$th or $[o_2\times100]$th percentiles of the bootstrap distribution, where $o_{1}=\Phi\left(2\widehat{z}_{0}+\Phi^{-1}(\alpha)\right) $ and $o_{2}=\Phi\left(2\widehat{z}_{0}+\Phi^{-1}(1-\alpha)\right)$. {Note that the bias-corrected confidence interval need not contain the point estimate if the finite-sample bias is large.}


\section{Specification Test of Location Invariance}
\label{sec:appx_test}

Given that our semiparametric locationally varying production model nests a more traditional fixed-parameter specification that implies locational invariance/homogeneity of the production function and the productivity evolution as a special case, we can formally discriminate between the two models to see if the data support our more flexible modeling approach. 

More concretely, we test the null hypothesis of a production model in which the technology is \textit{common} to firms across all locations:
\begin{equation}\label{eq:prodfn_hicks_cd_noS}
	\ln F_{|S_i}(\cdot) = \ln F(\cdot) =\beta_K k_{it}+\beta_L l_{it}+\beta_M m_{it} ,
\end{equation}
and firm productivity evolves according to a location-homogeneous first-order Markov process with  
\begin{equation}\label{eq:productivity_hicks_lawsp_noS}
	h^{\omega}_{|S_i}(\cdot ) = h^{\omega}(\cdot ) = \rho_0+ \rho_1 \omega_{it-1}+\rho_2G_{it-1}.
\end{equation} 

The location-invariant fixed-coefficient analogue of our model under the null of locational homogeneity is therefore given by 
\begin{align}
	y_{it} = \beta_K k_{it}+\beta_Ll_{it}+\beta_Mm_{it} +  \rho_0+ \rho_1 \omega_{it-1}+\rho_2G_{it-1} + \zeta_{it}+ \eta_{it}, \label{eq:prodfn_hicks_cd3_noS}
\end{align}
which we test against our semiparametric varying-coefficient alternative in \eqref{eq:prodfn_hicks_cd3}. This is, essentially, the test of overall relevancy of $S_i$. 

To test this hypothesis, we use \citeauthor{ullah1985}'s (1985) nonparametric goodness-of-fit test based on the comparison of the restricted (under H$_0$) and unrestricted (under H$_1$) models. First, let the estimator under H$_0$  be denoted by ``tilde'' whereas the estimator under H$_1$ be denoted by ``hat.'' Then, the residual-based test statistic is $T_n=(RSS_0-RSS_1)/RSS_1$, where $RSS_{0}=\sum_i\sum_t(\widetilde{ \zeta_{it}+ \eta_{it}})^2$ and $RSS_{1}=\sum_i\sum_t(\widehat{ \zeta_{it}+ \eta_{it}})^2$ are respectively the second-step residual sum of squares under the (restricted parametric) null and the (unrestricted semiparametric) alternative.\footnote{We use the \textit{second}-step residuals because they already incorporate information about the first-step estimation by virtue of sequential construction.} Intuitively, the test statistic is expected to converge to zero under the null and is positive under the alternative; hence the test is one-sided. To approximate the null distribution of $T_n$, we use wild panel-data block-bootstrap by resampling residuals from the model under the null. 

To approximate the null distribution of $T_n$, we use wild panel-data block-bootstrap by resampling residuals from model under the null. {The algorithm builds on that described in Appendix \ref{sec:appx_inference}.} 
\begin{enumerate}\itemsep 0pt
	\item Using the original data, compute the two steps of both the locationally-invariant (under H$_0$) and locationally varying (under H$_1$) models. Denote the estimates from the restricted model as $\big[\widetilde{\beta}_M,\widetilde{\theta},\widetilde{\Theta}'\big]'$ and the estimates from the unrestricted alternative as $\big[\widehat{\beta}_M(S_i),\widehat{\theta},\widehat{\Theta}(S_i)'\big]'$ for all $i=1,\dots,n$. Let the (negative of) first-step residuals under the null be $\{\widetilde{\eta}_{it}\}$ and those under the alternative be $\{\widehat{\eta}_{it}\}$. Also, obtain the second-step residuals under the null $\{\widetilde{ \zeta_{it}+ \eta_{it}}\}$ and under the alternative $\{\widehat{ \zeta_{it}+ \eta_{it}}\}$. Use the latter to compute the test statistic $T_n$. 
	
	\item Generate bootstrap weights $\xi_{i}^b$ for all cross-sectional units $i=1,\dots,n$ from the \citet{mammen1993} two-point mass distribution in \eqref{eq:two_point}. Next, for each observation $(i,t)$ with $i=1,\dots,n$ and $t=1,\dots,T$, generate a new bootstrap first-step disturbance $\eta_{it}^b = \xi_{i}^b\times\widetilde{\eta}_{it}$ and a new bootstrap second-step disturbance $(\zeta_{it}+ \eta_{it})^b=\xi_{i}^b\times(\widetilde{ \zeta_{it}+ \eta_{it}})$. When constructing these bootstrap disturbances, use re-centered residuals.
	
	\item Generate a new bootstrap first-step outcome variable based on the specification under H$_0$. From the first step, we have $v_{it}^b=\ln [\widetilde{\beta}_M\widetilde{\theta}]-\eta_{it}^b$ for all $i=1,\dots,n$ and $t=1,\dots,T$. 
	
	\item Recompute the first step of both the locationally-invariant and locationally varying models using $\{v_{it}^b\}$ in place of $\{v_{it}\}$ and denote the obtained parameter estimates as $\big[\widetilde{\beta}_M^b,\widetilde{\theta}^b\big]$ under the null and as $\big[\widehat{\beta}_M^b(S_i),\widehat{\theta}^b\big]$ under the alternative. 
	
	\item Generate a new bootstrap second-step outcome variable based on the specification under H$_0$. From the second step, we have $y_{it}^{*b}=\widetilde{\beta}_K k_{it}+\widetilde{\beta}_Ll_{it}+ \widetilde{\rho}^{\omega}_0+ \widetilde{\rho}^{\omega}_1 \Big[\widetilde{\nu}^{*}_{it-1}-\widetilde{\beta}_K k_{it-1}-\widetilde{\beta}_Ll_{it-1}\Big]+\widetilde{\rho}^{\omega}_2G_{it-1}+(\zeta_{it}+ \eta_{it})^b$ for all $i=1,\dots,n$ and $t=1,\dots,T$, where $\widetilde{\nu}^{*}_{it-1}=\ln[ P_{t-1}^M/P_{t-1}^Y]-\ln [\widetilde{\beta}_M\widetilde{\theta}]+[1-\widetilde{\beta}_M]m_{it-1}$ is constructed using the original parameter estimates. 
	
	\item Recompute the second step of both models using $\{y_{it}^{*b}\}$ in place of $\{y_{it}^{*}\}$. When recomputing the models, also use $\widetilde{\nu}^{*b}_{it-1}$ in place of $\widetilde{\nu}^{*}_{it-1}$ for the restricted model and $\widehat{\nu}^{*b}_{it-1}$ in place of $\widehat{\nu}^{*}_{it-1}$ for the unrestricted model, where $\widetilde{\nu}^{*b}_{it-1}=\ln[ P_{t-1}^M/P_{t-1}^Y]-\ln [\widetilde{\beta}_M^b\widetilde{\theta}^b]+[1-\widetilde{\beta}_M^b]m_{it-1}$ and $\widehat{\nu}^{*b}_{it-1}=\ln[ P_{t-1}^M/P_{t-1}^Y]-\ln [\widehat{\beta}_M^b(S_i)\widehat{\theta}^b]+[1-\widehat{\beta}_M^b(S_i)]m_{it-1}$ are constructed using the bootstrap parameter estimates. Denote the obtained parameter estimates as  $\widetilde{\Theta}^b=[\widetilde{\beta}_K^b,\widetilde{\beta}_L^b,\widetilde{\rho}^{\omega b}_0,\widetilde{\rho}^{\omega b}_1, \widetilde{\rho}^{\omega b}_2]'$ and $\widehat{\Theta}^b(S_i)=[\widehat{\beta}_K^b(S_i),\widehat{\beta}_L^b(S_i),$ $\widehat{\rho}^{\omega b}_0(S_i),\widehat{\rho}^{\omega b}_1(S_i), \widehat{\rho}^{\omega b}_2(S_i)]'$.
	
	\item Recompute the second-step bootstrap residuals under the null $\{\widetilde{ \zeta_{it}+ \eta_{it}}\}^b=y_{it} - \widetilde{\beta}_K^b k_{it}-\widetilde{\beta}_L^bl_{it}-\widetilde{\beta}_M^bm_{it} - \widetilde{\rho}^{\omega b}_0- \widetilde{\rho}^{\omega b}_1 \Big[\widehat{\nu}^{*b}_{it-1}-\widetilde{\beta}_K^b k_{it-1}-\widetilde{\beta}_L^bl_{it-1}\Big]-\widetilde{\rho}^{\omega b}_2G_{it-1}$ and under the alternative $\{\widehat{ \zeta_{it}+ \eta_{it}}\}^b=y_{it} - \widehat{\beta}_K^b(S_i) k_{it}-\widehat{\beta}_L^b(S_i)l_{it}-\widehat{\beta}_M^b(S_i)m_{it} - \widehat{\rho}^{\omega b}_0(S_i)- \widehat{\rho}^{\omega b}_1(S_i) \Big[\widehat{\nu}^{*b}_{it-1}-\widehat{\beta}_K^b(S_i) k_{it-1}-\widehat{\beta}_L^b(S_i)l_{it-1}\Big]-\widehat{\rho}^{\omega b}_2(S_i)G_{it-1}$. Use these to compute the bootstrap test statistic $T_n^b$. 
	
	\item Repeat steps 2 through 7 of the algorithm $B$ times. 
\end{enumerate} 

Use the empirical distribution of $ B+1$ bootstrap statistics $\left\{T_n^b\right\}$, where the first bootstrap replica is the test statistic $T_n$ calculated from the original data in Step 1, to obtain the $p$-value as $\sum_b\mathbbm{1}\left\{T^{b}_n\ge T_n\right\}/(B+1)$. {In our empirical application, the number of bootstrap iteration is set to $B=999$.

\medskip

\noindent\textbf{Estimating the Location-Invariant Model.}\textemdash 
The location-invariant model specified in \eqref{eq:prodfn_hicks_cd_noS}--\eqref{eq:productivity_hicks_lawsp_noS} is fully parametric and a special case of our locationally-varying model when $S_i=S_0$ for all $i$. It is therefore can be estimated following our methodology in \eqref{eq:fst_est2_pre}--\eqref{eq:sst_est} but by letting the adaptive bandwidths in both steps [$R_{h_1}(s)$ and $R_{h_2}(s)$] diverge to $\infty$ which would, in effect, obviate the need to locally weight the data because all kernels will be the same. We can then set $\mathcal{K}_{{h}_1}(S_i,s)=\mathcal{K}_{{h}_2}(S_i,s)=1$ for all $i$.

When the production technology and the productivity process are global and do not vary across locations as described in \eqref{eq:prodfn_hicks_cd_noS}--\eqref{eq:productivity_hicks_lawsp_noS}, the location-invariant analogue of the first-step material share equation in \eqref{eq:fst} is given by
\begin{equation}\label{eq:fst_locinv}
v_{it} = \ln [\beta_M\theta] - \eta_{it} ,
\end{equation}
and that of the second-step proxied production function in \eqref{eq:sst} is given by 
\begin{align}\label{eq:sst_locinv}
y_{it}^* &= \beta_K k_{it}+\beta_Ll_{it}+ \rho_0+ \rho_1 \Big[\nu^*_{it-1}-\beta_K k_{it-1}-\beta_Ll_{it-1}\Big]+\rho_2G_{it-1} + \zeta_{it}+ \eta_{it} ,
\end{align}
where $\nu^*_{it-1}=\ln[ P_{t-1}^M/P_{t-1}^Y]-\ln [\beta_M\theta]+[1-\beta_M]m_{it-1}$.

To estimate the material elasticity, denoting the unknown $\ln [\beta_M\theta]$ as some constant $b_M$, we have the following counterpart of the estimator in \eqref{eq:fst_est2_pre} which is just a sample mean:
\begin{equation}\label{eq:fst_est2_pre_locinv}
\widehat{b}_M = \frac{1}{nT}\sum_i\sum_t v_{it}.
\end{equation}
With it, we obtain the counterpart of \eqref{eq:fst_est2} that estimates $\beta_M$:
\begin{align}\label{eq:fst_est2_locinv}
\widehat{\beta}_M &= nT\exp\left\{ \widehat{b}_M \right\}\Big/ \sum_i\sum_t\exp\left\{ \widehat{b}_M-v_{it}\right\} \notag \\
&= nT\exp\left\{ \frac{1}{nT}\sum_i\sum_t v_{it} \right\}\Big/ \sum_i\sum_t\exp\left\{ \frac{1}{nT}\sum_i\sum_t v_{it}-v_{it}\right\}.
\end{align}

Using $\widehat{y}_{it}^*\equiv y_{it} - \widehat{\beta}_M m_{it}$ and $\widehat{\nu}^*_{it-1}=\ln[ P_{t-1}^M/P_{t-1}^Y]-\ln [\widehat{\beta}_M\theta]+[1-\widehat{\beta}_M]m_{it-1}$, we arrive at the location-invariant counterpart of the second-step estimator in \eqref{eq:sst_est2}:
\begin{align}\label{eq:sst_est2_locinv}
\widehat{\Theta} = \arg\min_{\Theta}\ \sum_i\sum_t &\ \Big( \widehat{y}_{it}^* - \beta_K k_{it}-\beta_Ll_{it} -
 \rho_0- \rho_1 \Big[\widehat{\nu}^*_{it-1}-\beta_K k_{it-1}-\beta_Ll_{it-1}\Big]+\rho_2G_{it-1} \Big)^2,
\end{align}
where $\Theta=[\beta_K,\beta_L,\rho_0,\rho_1,\rho_2]'$ and which can be estimated via the usual nonlinear least squares. }

{For inference, we follow the same bootstrap steps as those for our main model in Appendix \ref{sec:appx_inference} except that the estimated location-invariant fixed coefficients are used in place of the locationally-varying coefficients.}


{\footnotesize \setlength{\bibsep}{-1pt}  \bibliography{LocalHeterobib} }


\end{document}